\documentclass[pdftex,twocolumn,epjc3]{svjour3}

\RequirePackage[T1]{fontenc}
\usepackage{amsmath}
\RequirePackage{graphicx}
\RequirePackage{mathptmx}      
\RequirePackage{flushend}
\RequirePackage[numbers,sort&compress]{natbib}
\RequirePackage[colorlinks,citecolor=blue,urlcolor=blue,linkcolor=blue]{hyperref}

\usepackage{color}
\definecolor{darkgreen}{rgb}{0,0.5,0}
\definecolor{darkblue}{rgb}{0,0,0.7}
\definecolor{darkred}{rgb}{0.5,0,0.0}
\definecolor{darkorange}{rgb}{0.8,0.4,0.0}

\usepackage{amsmath}
\usepackage{amssymb}
\usepackage{xspace}
\usepackage{subfig}
\usepackage{comment}
\journalname{Eur. Phys. J. C}
\DeclareMathOperator{\De}{d}
\newcommand{\de}{\De\!}
\newcommand{\as}{\alpha_{\text{S}}}
\newcommand{\tg}{\theta_g}

\newcommand{\zcut}{z_c}
\newcommand{\SD}{Soft Drop\xspace}
\newcommand{\vi}{\ensuremath{\vartheta_{g,i}^2}}

\newcommand{\cf}{C_{\text F}}
\newcommand{\ca}{C_{\text A}}

\newcommand{\sherpa}{S\protect\scalebox{0.8}{HERPA}\xspace}
\newcommand{\pythia}{P\protect\scalebox{0.8}{YTHIA}8.3\xspace}
\newcommand{\herwig}{H\protect\scalebox{0.8}{ERWIG}\xspace}
\newcommand{\rivet}{R\protect\scalebox{0.8}{IVET}\xspace}
\newcommand{\fastjet}{F\protect\scalebox{0.8}{AST}J\protect\scalebox{0.8}{ET}\xspace}

\usepackage{cprotect}

\begin{document}

\title{On heavy-flavour jets with Soft Drop}
\author{
Simone Caletti\thanksref{e1,addr1}
\and
Andrea Ghira\thanksref{e2,addr2}
\and
Simone Marzani\thanksref{e3,addr2}
}

\thankstext{e1}{e-mail: scaletti@phys.ethz.ch}
\thankstext{e2}{e-mail: andrea.ghira@ge.infn.it}
\thankstext{e3}{e-mail: simone.marzani@ge.infn.it}

\institute{Institute for Theoretical Physics, ETH, CH-8093 Z\"urich, Switzerland\label{addr1}
\and
Dipartimento di Fisica, Universit\`a di Genova and INFN, Sezione di Genova, Via Dodecaneso 33, 16146, Italy\label{addr2}
}

\date{}


\maketitle

\begin{abstract}
We study hadronic jets that are tagged as heavy-flavoured, i.e.\ they contain either beauty or charm. In particular, we consider heavy-flavour jets that have been groomed with the Soft Drop algorithm. 
In order to achieve a deeper understanding of these objects,  we apply resummed perturbation theory to jets initiated by a massive quark and we perform analytic calculations for two variables that characterise Soft Drop jets, namely the opening angle and the momentum fraction of the splitting that passes Soft Drop. We compare our findings to Monte Carlo simulations. 
Furthermore, we investigate the correlation between the Soft Drop energy fraction and alternative observables that aim to probe heavy-quark fragmentation functions. 
\end{abstract}


\section{Introduction and motivation}\label{sec:intro}

Beauty ($b$) and charm ($c$) quarks are often considered ``heavy flavours'' because their masses are above the proton mass, $m_b=4.2$~GeV and $m_c=1.3$~GeV, respectively. However, their mass is not so large, compared to the typical scale of hadron formation, $\Lambda\simeq 1$~GeV, so hadronisation occurs. 
This does not happen for the top quark, which has a mass 170 times bigger than the proton mass. This value is so large that its lifetime is smaller than the hadronisation scale.

Heavy flavours constitute a window on two mechanisms that provide ordinary matter with mass: electroweak symmetry breaking and the binding energy of strong interactions. On the one hand, they play a crucial role in studies of the Higgs boson and, on the other hand, they constitute a bridge between perturbative and non-perturbative Quantum Chromo Dynamics (QCD).
For these reasons, they have been (and still are) the subject of detailed theoretical and experimental studies. 

In this work, we focus on processes in which heavy quarks are produced. 
From a theoretical point of view, calculations for identified heavy flavours can be performed essentially because the quark mass sets a perturbative scale for the running coupling and, at the same time, removes collinear singularities. 
From an experimental viewpoint, the lifetime of $B$ (or $D$) hadrons is long enough so that their decay happens away from the interaction point. Dedicated $b$- and $c$-tagging techniques that exploit this property to identify $B$ and $D$ hadrons, or $b$ and $c$ jets, are widely used in collider experiments.

Flavour physics has been studied for decades, both in the quark and lepton sectors. 
However, the recent development of Infra-Red and Collinear (IRC) safe flavour-jet algorithms~\cite{Banfi:2006hf,Caletti:2022hnc,Czakon:2022wam,Gauld:2022lem,Caola:2023wpj} opens up the possibility of setting up a yet-unexplored flavour physics program that exploits jets and their substructure at the Large Hadron Collider (LHC).

Resummed calculations for jets initiated by heavy quarks have been first performed in the context of studies focussing on $B$ decays~\cite{Aglietti:2006wh,Aglietti:2007bp,Aglietti:2008xn,Aglietti:2022rcm}
and top jets~\cite{Fleming:2007qr,Fleming:2007xt,Bachu:2020nqn,Jain:2008gb,Hoang:2019fze,Bris:2020uyb}, exploiting effective field theories.
However, to the best of our knowledge, there exists only a handful of studies that exploit modern jet substructure techniques to study heavy flavours~\cite{Maltoni:2016ays, Lee:2019lge,Llorente:2014bha,Li:2017wwc,Li:2021gjw,Craft:2022kdo,Cunqueiro:2022svx,Fedkevych:2022mid}. 
The main goal of these studies is to investigate the so-called dead-cone effect~\cite{Dokshitzer:1991fd,Dokshitzer:1995ev}, i.e.\ the fact that colour radiation around heavy quarks is suppressed. Remarkably, this effect has been recently measured by the ALICE collaboration at the LHC \cite{ALICE:2021aqk}.

In contrast, there exists a rather extensive literature dedicated to studying the properties of a reconstructed $B$ (or $D$) hadron, such as its energy or its transverse momentum. 
The theoretical description of these observables is usually based on heavy-quark fragmentation functions, which can be computed in perturbative QCD. State-of-the-art predictions include the resummation of different classes of logarithmic corrections, e.g.\ mass logarithms and end-point logarithms, see e.g.~\cite{Mele:1990cw,Mele:1990yq,Melnikov:2004bm,Mitov:2004du,Cacciari:2001cw,Fickinger:2016rfd,Maltoni:2022bpy,Czakon:2022pyz}. In this context, two of us have recently developed a theoretical framework to consistently resum both mass and soft logarithms~\cite{Gaggero:2022hmv,Ghira:2023bxr}. In this study, we will heavily rely on these results and we will apply them to jet-based observables. This way, we will be able to obtain theoretical predictions that resum both the logarithms of the observable we want to study and the logarithms of the ratio of the heavy-quark mass to the jet transverse momentum.

In this study, we investigate the possibility of exploiting a widely used jet substructure technique, namely Soft Drop~\cite{Larkoski:2014wba}, to study heavy-flavour jets. 
The Soft Drop algorithm has been extensively studied and Soft Drop jets are routinely used in experimental analyses, both in the context of measurements and searches for new physics.
In particular, we consider heavy-flavour-tagged jets and focus on the angular separation ($\theta_g$) and momentum fraction ($z_g$) of the first splitting that passes Soft Drop. The former directly measures the angular resolution of the groomed jet and, therefore, it gives us direct access to the dead cone. The latter, instead, allows us to probe the heavy-quark splitting function~\cite{Larkoski:2015lea,Ilten:2017rbd}. 

These observables have been recently measured on $c$-jets by the ALICE collaboration~\cite{ALICE:2022phr} and our study constitutes the first step towards a first-principle description of the data. A direct comparison to the ALICE data, however, goes beyond the scope of this paper, as it would require matching the resummed expressions to fixed-order matrix elements, as well as accounting for the fact that ALICE reconstructs $c$-jets from exclusive $D$-meson decays.

The paper is organised as follows. We start by reviewing the Soft Drop algorithm in Sec.~\ref{sec:softdrop}. In Sec.~\ref{sec:theta_g} we  perform the next-to-leading logarithmic (NLL) resummation of the $\theta_g$ distribution, including heavy-quark mass effects and compare it Monte Carlo parton-shower simulations. In Sec.~\ref{sec:z_g}
we study the $z_g$ distribution, both from first-principle and in Monte Carlo simulations and study its correlation to a standard momentum fraction variable used in the context of fragmentation functions. Finally, we draw our conclusions in Sec.~\ref{sec:conclusions}. Details of the calculations are collected in the Appendices.

\section{Soft Drop}\label{sec:softdrop}

 Soft Drop \cite{Larkoski:2014wba} is a grooming algorithm that recursively removes soft-wide angle constituents from a jet. The Soft Drop procedure starts by re-clustering a given jet (typically an anti-$k_t$~\cite{Cacciari:2008gp} jet) with radius $R_0$ and transverse momentum $p_t$ with the Cambridge-Aachen (C/A) algorithm~\cite{Wobisch:1998wt, Dokshitzer:1997in}. Soft Drop then parses the resulting angular-ordered branching history, grooming away the softer branch, until a branch that satisfies the condition 
\begin{align}
    \label{eq:sdcrit}
\frac{\min \left( p_{t1}, p_{t2}\right)}{p_{t1} + p_{t2}} & > z_c\left( \frac{\Delta_{12}}{R_0}\right)^\beta, \\ \nonumber &\Delta_{12}=\sqrt{(y_1-y_2)^2+(\phi_1-\phi_2)^2},
\end{align}
is found.
In the expression above, $1$ and $2$ denote the branches at a given step in the clustering, $p_{ti}$ are the corresponding transverse momenta, and $\Delta_{12}$ is their rapidity-azimuth separation.  
If the condition above is never satisfied, we can either remove the jet from
consideration (“tagging mode”) or leave it as the final Soft Drop jet (“grooming mode”).

The kinematics of the first branch that satisfies (\ref{eq:sdcrit}) defines the groomed jet radius $\theta_g$ and the groomed momentum sharing $z_g$:
\begin{equation}
\theta_g = \frac{\Delta_{12}}{R_0}, \qquad z_g = \frac{\min \left(p_{t1}, p_{t2}\right)}{p_{t1} + p_{t2}}.
\end{equation}
The
$\theta_g$ distribution is IRC safe and it was first studied, for light jets, in~\cite{Larkoski:2014wba} and then resummed to next-to-leading logarithmic (NLL) accuracy in~\cite{Kang:2019prh}. The momentum sharing $z_g$ is IRC safe for $\beta<0$ but only Sudakov safe for $\beta \ge0$~\cite{Larkoski:2015lea}. The NLL calculation of the $z_g$ distribution was performed in~\cite{Cal:2021fla}.

\section{The $\theta_g$ distribution}\label{sec:theta_g}
In this section, we describe the resummation of the $\theta_g$ distribution for a jet originated by a massive quark. 
We follow the general strategy described in Ref.~\cite{Larkoski:2014wba} and we, therefore, consider the resummation of the cumulative distribution, normalised to the Born cross section,
\begin{equation}
    \Sigma(\theta_g^2)=\frac{1}{\sigma_0}\int_0^{\theta_g^2}\frac{\de \sigma}{\de {\theta_g'}^2} \de {\theta_g'}^2,
\end{equation}
exploiting Lund diagrams~\cite{Andersson:1988gp}.

\subsection{Lund plane geography}\label{sec:lund}
\begin{figure*} 
\begin{center}
\includegraphics[page=1,width=0.49\textwidth]{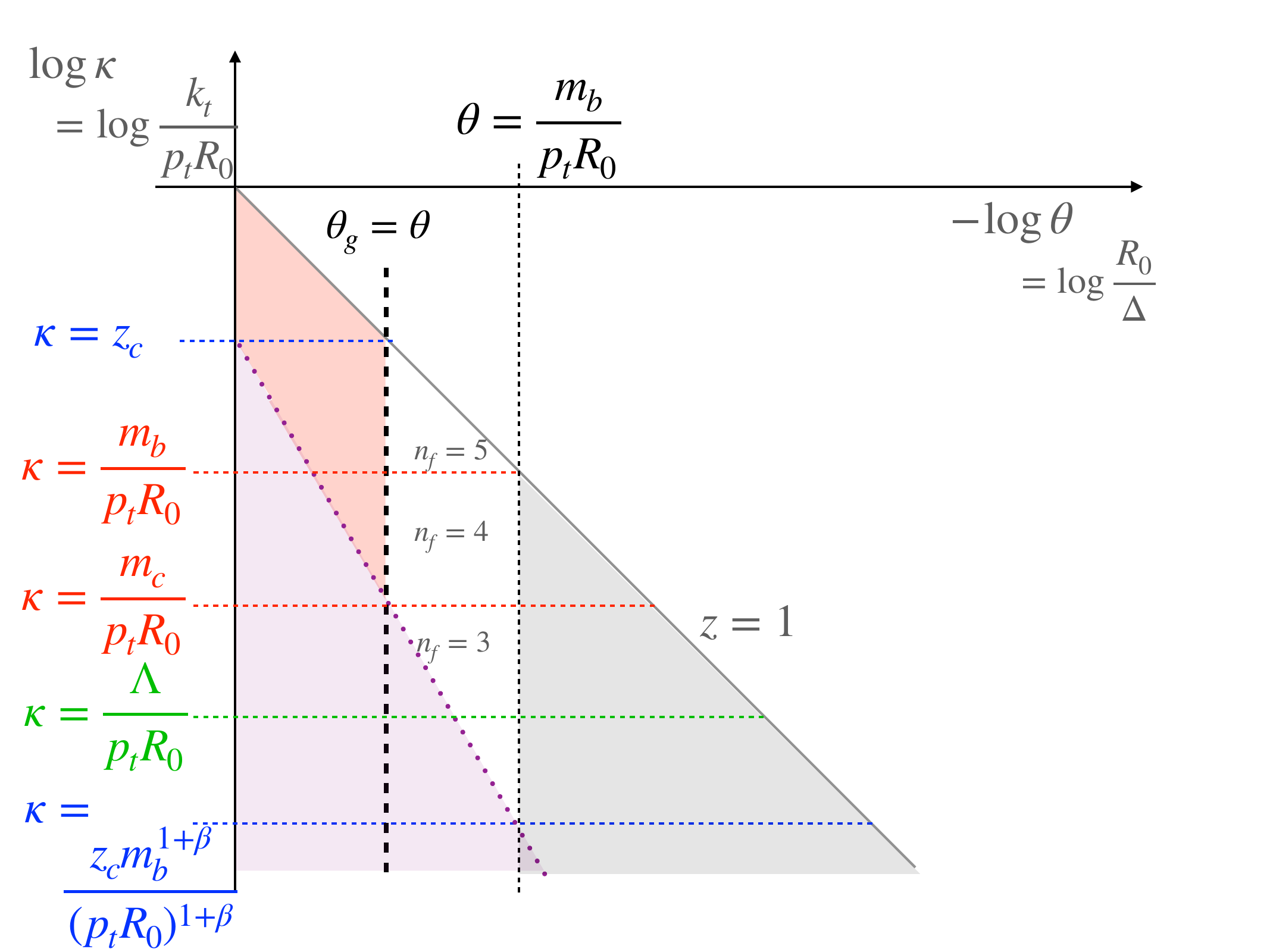}
\includegraphics[page=2,width=0.49\textwidth]{figures/lund.pdf}
    \end{center}
\caption{Lund plane representation of the soft and quasi-collinear phase space for emissions off a $b$ quark, on the left, and off a $c$ quark, on the right. The dead-cone region is indicated in grey. Soft Drop is applied and the groomed-away region appears in light purple. The vertical dashed line in black indicates a measurement of the groomed radius $\theta_g$ and the corresponding area in pink is the vetoed region, which gives rise to the Sudakov form factor. The horizontal red lines indicate the boundaries between different flavour numbers for the running coupling, while the green one marks the boundary of the non-perturbative region.}
\label{fig:lundplane}
\end{figure*}
Lund diagrams are a useful way to represent the available phase space for the emission of soft and collinear gluons off a hard dipole. 
 The Lund plane is usually represented in terms of pairs of the logarithmic variables, so that, in the soft and collinear limit, equal areas correspond to equal emission probabilities. 
We choose to draw Lund planes as depicted in Fig.~\ref{fig:lundplane}.
On the horizontal axis, we show the logarithm of the emission angle with respect to the jet axis, as measured in the rapidity-azimuth plane, normalised to the jet radius, i.e.\ $\theta=\frac{\Delta}{R_0}$. On the vertical axis, we show the emission's transverse momentum with respect to the jet axis, normalised to the hard scale of the process, $\kappa=\frac{k_t}{p_t R_0}$.
 This way, lines of constant $\theta_g$, in the soft and collinear limits, are represented by straight vertical lines, as shown in Fig.~\ref{fig:lundplane}, in dashed black.
 The Soft Drop condition $z=z_c \theta^\beta$ sets the  dotted line in purple. In this study, we consider $\beta\ge0$. The region of phase space below this line (shaded in purple) is groomed away.

Recently, two of us have achieved a generalisation of the Lund-plane resummation formalism that includes quark-mass effects~\cite{Ghira:2023bxr}.
The Lund plane on the left-hand panel of Fig.~\ref{fig:lundplane} is for a $b$-quark line, while the one on the right is for a $c$-quark.
When considering heavy-quark Lund planes, two important differences with respect to the massless case arise. First, the presence of the vertical short-dashed line, in black, that represents the beginning of the dead cone at $\theta=\frac{m_b}{p_t R_0}$, i.e.\ $\Delta=\frac{m_b}{p_t}$, for the left-hand figure, and at $\theta=\frac{m_c}{p_t R_0}$, for the right-hand one. 
The shaded area in grey is the dead-cone region. Here, collinear emissions do not give rise to any logarithmic enhancement, leading to a suppression of QCD activity in this region. 
Second, the presence of heavy-quark thresholds, which are relevant when considering the running of the strong coupling. In Fig.~\ref{fig:lundplane}, they are represented by horizontal dotted lines in red. 
The lines at $\kappa=\frac{m_{b,c}}{p_t R_0}$ correspond to $k_t=m_{b,c}$ and therefore mark the boundaries between two regions with different numbers of active flavours: $n_f=5$ above and $n_f=4$ below, and $n_f=4$ above and $n_f=3$ below, for $k_t=m_b$ and $k_t=m_c$, respectively. 
We also show (in green) the line $\kappa=\frac{\Lambda}{p_t R_0}$ that marks the $k_t=\Lambda\simeq 1$~GeV boundary between the perturbative and the non-perturbative regions. For the region below this line, we need a prescription to regulate the Landau pole of the strong coupling. Details are given in~\ref{app:rc}.
Finally, we show dotted lines in blue, at  $\kappa=z_c$ and at $\kappa=z_c \left(\frac{m_{b,c}}{p_t R_0}\right)^{1+\beta}$, which originate from the Soft Drop condition and the dead-cone boundary.

When performing actual calculations, one needs to establish the hierarchy between these scales. 
Let us consider, as an example the case of $b$-jets, with relatively high-$p_t$. Because Soft Drop is usually employed with $z_c=0.1$, we have that $z_c> \frac{m_b}{p_t R_0}$ if $p_t R_0> 42$~GeV.
Ideally, we would like to have $z_c \left(\frac{m_{b}}{p_tR_0} \right)^{1+\beta}>\frac{\Lambda}{p_t R_0}$. This results in an upper bound for the hard scale $(p_t R_0)^\beta<z_c \frac{m_{b}^{1+\beta}}{\Lambda}$. However, this inequality is hardly satisfied if we consider commonly used values of the Soft Drop parameters, e.g. $z_c=0.1$ and $\beta=0$. Thus, we do expect non-perturbative contributions to affect the dead-cone region and we will assess their size using Monte Carlo simulations. 
In Fig.~\ref{fig:lundplane}, we show the Lund plane for this hierarchy of scales, on the left for $b$-jets and on the right for $c$-jets.
A discussion of all the other possible hierarchies and regions can be found in~\ref{app:regions}.

\subsection{Recap of the massless calculation}\label{sec:theta_g_massless}
Before computing the $\theta_g$ resummed distribution for a jet initiated by a massive quark, let us briefly recall the structure of the resummation in the case of massless partons. The all-order cumulative distribution can be written as the product of two contributions
\begin{equation}\label{eq:masslessNLL}
    \Sigma^\text{NLL}(\theta_g^2)=S(\theta_g^2)\, e^{-R(\theta_g^2)}.
\end{equation}
The function $R$ is the Sudakov exponent, or radiator, which accounts for angular-ordered collinear emissions. It was first computed in~\cite{Larkoski:2014wba} and, for the case of a quark jet, reads
\begin{align}\label{eq:sudakov}
    R(\theta_g^2) = \int_{\tg^2}^1 \frac{\de\theta^2}{\theta^2}\int_0^1 \de z \,P_{gq}(z)\frac{\as^\text{CMW}(k_t^2)}{2\pi}\Theta(z-z_c \theta^\beta)\,,
\end{align}
where $\theta$ is the emission angle, normalised to the jet radius, $k_t= z \theta p_t R_0$ and the massless $q\to q g$ (timelike) splitting function is
\begin{equation}\label{eq:Pgq-massless}
    P_{gq}(z)=\cf \frac{1+(1-z)^2}{z}.
\end{equation}
In order to achieve NLL accuracy, the running coupling must be evaluated at two loops in the so-called CMW scheme~\cite{Catani:1990rr}.
Henceforth, we will also work in the small-$z_c$ limit. Consequently, even in the case $\beta=0$, we will ignore flavour-changing contributions~\cite{Dasgupta:2013ihk,Marzani:2017mva}.
The Sudakov form factor represents the no-emission probability and corresponds to the area shaded in pink on the Lund planes of Fig.~\ref{fig:lundplane}.

This rather simple picture essentially arises from two facts. First, for the $\theta_g$ distribution, independent emissions, i.e.\ $\as^n \cf^n$ contributions, exponentiate with no corrections due to multiple emissions~\cite{Larkoski:2014wba}. Second, all non-Abelian contributions are unresolved and therefore captured at NLL by the running coupling in the CMW scheme. 
However, as it was pointed out in~\cite{Kang:2019prh}, the above description is not complete at NLL and the resummed expression must be supplemented with a correction factor, $S$ in Eq.~(\ref{eq:masslessNLL}). 
Indeed simple exponentiation is broken by single logarithmic contributions that arise from two competing mechanisms. 
Let us start with the correction to the independent emission contribution. The C/A algorithm, which is used as the first step of the Soft Drop procedure, can cluster two soft gluons together first, if they are the closest pair in angle, instead of clustering each of them with the hard quark. This happens when $\theta_{12}<\min(\theta_1,\theta_2)$.  This effect introduces single-logarithmic corrections that are usually referred to as clustering logarithms~\cite{Banfi:2005gj,Delenda:2006nf,Delenda:2012mm,Banfi:2010pa}.
On the other hand, the clustering condition also plays a role in the correlated emission contribution. 
The C/A algorithm can resolve a soft gluon splitting, giving rise to a non-global contribution~\cite{Dasgupta:2001sh,Appleby:2002ke} that spoils the CMW picture. This is also a single logarithmic contribution that can happen if the two soft gluons are not the closest pair in angle, i.e.\ $\theta_{12}>\min(\theta_1,\theta_2)$.

Both contributions start at $\mathcal{O}(\as^2)$ and can be computed using the expression of the squared matrix element for the emission of two strongly-ordered soft gluons with momenta $k_1$, $k_2$ off a hard fermionic dipole with momenta $p_a$, $p_b$~\cite{Catani:1983bz,Dokshitzer:1992ip}
\begin{align} \label{eq:double-soft-massless}
    W &= 2 \cf w_{ab,1}\left[\ca w_{a1,2}+\ca w_{b1,2}+\left(2 \cf -\ca\right)w_{ab,2}\right]\nonumber\\
    &= 4\cf^2 \,   w_{ab,1} w_{ab,2} + 2 \cf \ca\, w_{ab,1}
    \left(w_{a1,2}+ w_{b1,2}-w_{ab,2}\right),
\end{align}
where the $\cf^2$ term describes the independent emission contribution, while the $\cf \ca$ one, the correlated one. We have introduced the eikonal factor
\begin{equation}\label{eq:w-massless}
    w_{ij,l}= \frac{p_i\cdot p_j}{p_i\cdot k_l \, p_j\cdot k_l}.
\end{equation}
Following~\cite{Kang:2019prh}, we consider the case in which the softer gluon ($k_2$) is emitted at a large angle and passes the Soft Drop condition, while the harder gluon ($k_1$) is emitted at an angle smaller than $\theta_g$ and so it is not subject to the Soft Drop condition. 
If we work in the small-angle limit, considering real and virtual contributions, we find
\begin{align}\label{eq:massless-clust-start}
    S&=1+\left(\frac{\as}{\pi}\right)^2\int_0^1\frac{\de z_1}{z_1}\int_0^1 \frac{\de z_2}{z_2}\int_0^{2 \pi}\frac{\de \phi}{2 \pi}\int_0^{\theta_g^2}  \frac{\de \theta_1^2}{\theta_1^2}\int_{\theta_g^2}^1\frac{\de \theta_2^2}{\theta_2^2} \nonumber \\
    &\times \Theta\left(z_2- z_c \theta_2^\beta\right)\Theta\left(z_1-z_2\right) 
    \Bigg[\cf^2\, \Theta\left(\theta_1-\theta_{12}\right) \nonumber \\ &-\cf \ca \frac{ \theta_1 \theta_2\; \cos \phi }{\theta_1^2+\theta_2^2-2 \theta_1 \theta_2 \cos \phi}\Theta\left(\theta_{12}-\theta_1\right)\Bigg]+\mathcal{O}(\as^3),
\end{align}
with $\theta_{12}^2=\theta_1^2+\theta_2^2-2 \theta_1 \theta_2 \cos \phi$. The integration above can be simplified by noting that, in order to capture the highest power of the logarithm of $\theta_g$, we can evaluate the momentum fraction integrals with lower limit $z_2= z_c \theta_2^\beta \simeq z_c \theta_g^\beta$. This way, the momentum fraction integrals decouple from the angular ones and we obtain
\begin{align}\label{eq:massless-clust-end}
    S&= 1+\frac{1}{2}\left(\frac{\as}{\pi}\right)^2 \log^2 \left(z_c \theta_g^\beta\right)\int_0^{2 \pi}\frac{\de \phi}{2 \pi}\int_0^{\theta_g^2}  \frac{\de \theta_1^2}{\theta_1^2}\int_{\theta_g^2}^1\frac{\de \theta_2^2}{\theta_2^2}
  \nonumber \\
    &\times\Bigg[\cf^2\, \Theta\left(2\theta_1\cos \phi-\theta_{2}\right) -\cf \ca \frac{ \theta_1 \theta_2\; \cos \phi }{\theta_1^2+\theta_2^2-2 \theta_1 \theta_2 \cos \phi} \nonumber \\ &\times\Theta\left(\theta_{2}- 2\theta_1\cos \phi\right)\Bigg]+\mathcal{O}\left( \as^2 \log \theta_g\right)\nonumber\\
    &
    =1+\left(\frac{\as}{\pi}\right)^2\log^2 \left(z_c \theta_g^\beta\right) \frac{\pi^2}{108}\left(\cf^2  - 4 \cf \ca\right),
\end{align}
The all-order resummation of these contributions has been performed numerically, in the large-$N_c$ limit~\cite{Kang:2019prh}. In this study, we have decided to limit ourselves to studying the impact of these corrections. Thus, we approximate the clustering factor $S$ by simply considering the exponentiation of the two-loop result:
\begin{equation}\label{eq:clustering-running-massless}
    S(\theta_g^2)= \exp{ \left[\frac{\cf^2  - 4 \cf \ca}{108}\left(\int_{z_c \theta_g^\beta}^1 \frac{\de z}{z}\, \as(z^2 p_t^2 R_0^2)\right)^2\right]},
\end{equation}
where the running of the strong coupling is taken at one loop. 

\subsection{The $\theta_g$ distribution for a heavy-flavour jet}
We now perform the resummation of the $\theta_g$ distribution for jets initiated by a heavy quark, considering both the case of a $b$-quark and a $c$-quark. 
Both the Sudakov contribution $R$ and the clustering correction function $S$ in Eq.~(\ref{eq:masslessNLL}) need to be reconsidered. We start with the calculation of the resummed Sudakov exponent.

It was realised long ago that squared QCD matrix elements with massive partons factorise in the so-called quasi-collinear limit~\cite{Catani:2000ef,Catani:2002hc}. In this approximation, both the transverse momentum $k_t$ of the emission with respect to the massive emitter, and the mass $m$ are small compared to the hard scale but they are considered of the same order. In this limit, the squared invariant amplitude for one-gluon emission takes the form
\begin{equation}\label{eq:quasi-coll-limit}
	|\mathcal{M}|^2\simeq 8\pi\as \frac{z(1-z)}{k_{t}^2+z^2 m_i^2} P_{gi}\left(z,k_{t}^2\right) 	|\mathcal{M}_0|^2, \quad i=b,c,
\end{equation}
where the massive splitting function for $i \to i g$ is
\begin{equation}\label{eq:Massive AP}
	P_{gi}(z,k_t^2)= \cf\left(\frac{1+(1-z)^2}{z}-\frac{2 z(1-z)m_i^2}{k_{t}^2+z^2 m_i^2}\right).
\end{equation}
Following the same steps as in~\cite{Ghira:2023bxr}, we write
\begin{align} \label{eq:massive-radiator-start}
      R_i(\theta_g^2,m_i^2,m_b^2,m_c^2)&= 
     \int_0^{p_t^2 R_0^2} \frac{\de k_t^2}{k_t^2+z^2 m_i^2}\int_0^1 \de z \,P_{gi}(z,k_t^2)
   \\ &\times
     \frac{\as^\text{CMW}(k_t^2)}{2\pi} \Theta(z-z_c  \theta^\beta) \Theta(\theta -\theta_g)\,,   \nonumber
\end{align}
where $\theta =\frac{k_t}{z p_t R_0}$.
Note that the mass dependence enters in two ways. First, for a jet initiated by a quark flavour $i=b,c$, the quasi-collinear phase-space and the massive splitting function depend on $m_i$. Second, regardless of the jet flavour, the running of the strong coupling may cross the $b$ and the $c$ thresholds, thus inducing a logarithmic dependence on the quark masses. 
In order to proceed further, it is convenient to change the integration variable in Eq.~(\ref{eq:massive-radiator-start}) from $k_t^2$ to $\theta^2$: 
\begin{align} \label{eq:massive-radiator-cntd}
      R_i(\theta_g^2,\theta_i^2,\xi_b,\xi_c)&= 
     \int_{\theta_g^2}^{1} \frac{\de \theta^2}{\theta^2+ \theta^2_i}\int_0^1 \de z \,P_{gi}\left(z, k_t^2\right)\frac{\as^\text{CMW}(k_t^2)}{2\pi} \nonumber\\ &\times \Theta(z-z_c  \theta^\beta) \,,
\end{align}
where now $k_t^2=z^2 \theta^2 p_t^2 R_0^2$,
 and we have introduced the following dimensionless variables
\begin{align}
    \xi_i=\frac{m_i^2}{p_t^2 R_0^2}, \quad  \theta_{i}^2 =\xi_i, \quad \text{with} \quad i=b,c.
\end{align} 
We stress again that the Sudakov exponent for either a high-$p_t$ $b$-jet or $c$-jet depends on both $\xi_b$ and $\xi_c$ because the running coupling crosses both quark thresholds. However, it depends only on the dimensional ratio $\theta_i^2$ of the corresponding flavour $i=b,c$. 

The mass-dependent shift in the denominator of Eq.~(\ref{eq:massive-radiator-cntd}) acts as an effective lower bound of a logarithmic angular integration.
This is the well-known dead-cone effect,~\cite{Dokshitzer:1991fd,Dokshitzer:1995ev} i.e.\ the fact that radiation off massive partons at angles below $m/p_t$ is not logarithmically enhanced. Therefore, we can further simplify our expression by shifting the angular integration variable to $\bar \theta^2=\theta^2+\theta^2_i$. Neglecting power corrections in the mass, we have
\begin{align} \label{eq:massive-radiator-cntd-2}
      &R_i(\theta_g^2,\theta_i^2,\xi_b,\xi_c)= 
     \int_{\vi}^{1} \frac{\de \bar \theta^2}{\bar \theta^2}\int_0^1 \de z \,P_{gi}\left(z, z^2 (\bar \theta^2-\theta_i^2)p_t^2 R_0^2\right)
     \nonumber\\ & \quad \quad \quad \;\times 
     \frac{\as^\text{CMW}(z^2 (\bar \theta^2-\theta_i^2)p_t^2 R_0^2)}{2\pi} \Theta(z-z_c  (\bar \theta^2-\theta_i^2)^\frac{\beta}{2}) \,,
\end{align}
with $\vi=\theta_g^2 +\theta^2_i$. Let us first note that
\begin{align}\label{eq:Massive-AP-bis}
	P_{gi}\left(z, z^2 (\bar \theta^2-\theta_i^2)p_t^2 R_0^2\right)&= \cf\left(\frac{1+(1-z)^2}{z}-\frac{2 \theta_i^2 (1-z)}{z \bar \theta^2}\right)
 \nonumber\\ &\equiv \mathcal{P}_{gi}(z,\bar \theta^2).
\end{align}
Furthermore, as discussed in detail in~\cite{Ghira:2023bxr}, the mass-dependent shift in the argument of the running coupling contributes at most to NNLL corrections and, therefore, it can be dropped at the accuracy we are working at.
A similar argument applies to the shift in the Soft Drop condition. This can be easily checked at fixed coupling. For instance, in the soft and collinear limits, we have
\begin{align}\label{eq:radiator-fc-shift}
      R_i^{(\text{f.c.})}&= \frac{\as \cf}{\pi}
     \int_{\vi}^{1} \frac{\de \bar \theta^2}{\bar \theta^2}\int_0^1 \frac{\de z}{z} \Theta(z-z_c  (\bar \theta^2-\theta_i^2)^\frac{\beta}{2})
     \nonumber\\
     &=\frac{\as \cf}{\pi}\Bigg \{ \log \vi \left( \log z_c+\frac{\beta}{4} \log \vi\right)\nonumber\\&-\frac{\beta}{2}\left[\text{Li}_2\left(-\frac{1}{\theta_i^2}\right)-\text{Li}_2\left(\frac{\theta_i^2}{\vi}\right)+\frac{1}{2}\log^2\theta_i^2+\frac{\pi^2}{6}\right]\Bigg\}\nonumber\\
     &= \frac{\as \cf}{\pi} \int_{\vi}^{1} \frac{\de \bar \theta^2}{\bar \theta^2}\int_0^1 \frac{\de z}{z} \Theta(z-z_c  \bar \theta^\beta)+ \text{NNLL},
\end{align}
where in the last step we have dropped constant terms in the $\theta_i \to 0$ or $\theta_g \to 0$ limits, which are NNLL.
Thus, the massive Sudakov exponent at NLL becomes 
\begin{align} \label{eq:massive-radiator-final}
      R_i(\theta_g^2,\theta_i^2,\xi_b,\xi_c)&= 
     \int_{\vi}^{1} \frac{\de \bar \theta^2}{\bar \theta^2}\int_{z_c  \bar \theta^\beta}^1 \de z \,\mathcal{P}_{gi}(z, \bar \theta^2)  \nonumber\\ & \quad \quad\quad\quad \times
     \frac{\as^\text{CMW}(z^2 \bar \theta^2 p_t^2 R_0^2)}{2\pi}.
\end{align}
Remarkably, with the sole exception of the mass-dependent term in the splitting function, this result, when seen as a function of $\vi$, has the same form of the massless case, Eq.~(\ref{eq:sudakov}). Consequently, we can represent the massive Sudakov form factor in Fig.~\ref{fig:lundplane} using the same Lund planes as in the massless case, provided that we interpret vertical lines as lines of constant $\vi$. This way, the dead-cone line acts as a phase-space boundary because $\vi \to \theta_i^2$ as $\theta_g \to 0$. 
The integrations in Eq.~(\ref{eq:massive-radiator-final}) are all straightforward but the presence of the Soft Drop condition, of the dead cone, and of mass thresholds force us to consider many cases. Details are given in the Appendices.

Next, we have to consider the clustering correction factor $S$. We adopt the same, approximate, strategy of the massless case. Namely, we consider the running-coupling exponentiation of the two-loop result. 
The square amplitude for the emission of two soft gluons off a massive quark-antiquark dipole was computed in \cite{Czakon:2011ve,Czakon:2014oma}. For our purposes, we are actually interested in a less general result, namely the case of the emission of two soft gluons that are strongly ordered in energies, e.g. $z_2\ll z_1$, off a single dipole. In this limit, the massive square matrix element takes the same form as Eq.~(\ref{eq:double-soft-massless}) but with the massless eikonal factor $w_{ij,k}$ substituted by the massive one:
\begin{equation}\label{eq:w-massive}
    w^m_{ij,l}= \frac{p_i\cdot p_j}{p_i\cdot k_l \, p_j\cdot k_l}-\frac{m_i^2}{2 p_i \cdot k_l}-\frac{m_j^2}{2 p_j \cdot k_l}.
\end{equation}
We choose $p_a$, see Eq.~(\ref{eq:double-soft-massless}), to be the momentum of the heavy quark and we work in the quasi-collinear limit with respect to its momentum direction. Therefore, we can set $p_b^2=0$. We then integrate the $\cf^2$ and $\cf \ca$ contributions with the same constraints as in Eq.~(\ref{eq:massless-clust-start}):
\begin{align}\label{eq:massive-clust-start}
    S_i&=1+\left(\frac{\as}{\pi}\right)^2\int_0^1\frac{\de z_1}{z_1}\int_0^1 \frac{\de z_2}{z_2}\int_0^{2 \pi}\frac{\de \phi}{2 \pi}\int_0^{\theta_g^2}  \frac{\de  \theta_1^2}{\bar \theta_1^2}\int_{\theta_g^2}^1\frac{\de \theta_2^2}{\bar \theta_2^2} \nonumber \\
    &\times \Theta\left(z_2- z_c \theta_2^\beta\right)\Theta\left(z_1-z_2\right) 
    \Bigg[\cf^2\, \Theta\left(\theta_1-\theta_{12}\right) 
    \nonumber\\& \times \left(1
    -\frac{\theta_i^2}{\bar \theta_1^2}
    -\frac{\theta_i^2}{\bar \theta_2^2}
    +\frac{\theta_i^4}{\bar \theta_1^2 \bar \theta_2^2}\right)
    -\cf \ca \frac{ \theta_1 \theta_2\; \cos \phi +\theta_i^2 }{\theta_1^2+\theta_2^2-2 \theta_1 \theta_2 \cos \phi}
    \nonumber \\ & \times \left(1-\frac{\theta_i^2}{\bar \theta_1^2} \right)
\Theta\left(\theta_{12}-\theta_1\right)\Bigg]+\mathcal{O}(\as^3), \quad i=b,c,
\end{align}
where, as before, $\bar \theta_{1}^2=\theta_{1}^2+\theta_i^2$ and $\bar \theta_{2}^2=\theta_{2}^2+\theta_i^2$. As in the massless case, in order to obtain the leading contribution we can substitute $\theta_2\simeq\theta_g$ in the Soft Drop condition.
Therefore, the momentum fraction integrals decouple from the angular ones and we obtain
\begin{align}\label{eq:massive-clust-end}
    S_i= 1+\left(\frac{\as}{\pi}\right)^2\log^2 z_c \theta_g^\beta &\Big[\cf^2\mathcal{F}_1(\theta_g^2,\theta_i^2)\nonumber \\&+ \cf \ca \mathcal{F}_2(\theta_g^2,\theta_i^2)\Big].
\end{align}
The Abelian contribution reads
\begin{align}\label{eq:F1}
\mathcal{F}_1(\theta_g^2,\theta_i^2)&=
\int^{\frac{\pi}{3}}_{0}\frac{\de \phi}{2\pi}\int^{4\cos^2{\phi}}_{1}\frac{x \de x}{(x+\kappa)^2}\Bigg[\log{(g(\phi))}
\nonumber \\
    &-\frac{\kappa }{1+\kappa}\left(g(\phi)-1\right)
    \Bigg],
\end{align}
with
\begin{equation}
g(\phi)=\frac{4(1+\kappa)\cos^2{\phi}}{x+4\kappa\cos^2{\phi}}, \quad \kappa= \frac{\theta_i^2}{\tg^2}.
\end{equation} 
For the non-Abelian case, instead, we find
\begin{align}\label{eq:F2}
        \mathcal{F}_{2}(\theta_g^2,\theta_i^2)&=-\int^{2\pi}_0 \frac{\de \phi}{2\pi}\int^{1}_0 \frac{\de y}{y(1+y^2 \kappa)} \int^{y}_0 \de t \frac{2t}{t^2+y^2 \kappa} 
        \\ \nonumber
        & \frac{t \cos{\phi}+y^2\kappa}{t^2-2t\cos{\phi}+1}
        \left[1-\frac{\kappa y^2}{t^2+\kappa y^2}\right]  \Theta(1-2 t \cos{\phi}). \nonumber
  \end{align}
  We evaluate $\mathcal{F}_1$ and $\mathcal{F}_2$ numerically. However, in the limit $\theta_g^2\gg \theta_i^2$ we are able to perform the integrals analytically and we obtain
\begin{align}
    \lim_{\theta_i^2/\theta_g^2 \to 0}\mathcal{F}_1(\theta_g^2,\theta_i^2)&=\frac{\pi^2}{108},\nonumber \\
     \lim_{\theta_i^2/\theta_g^2 \to 0}\mathcal{F}_2(\theta_g^2,\theta_i^2)&=-\frac{\pi^2}{27},
\end{align}
so that the massless result Eq.~(\ref{eq:massless-clust-end}) is recovered.
Remarkably, in the opposite limit, we find
\begin{align}
    \lim_{\theta_g^2/\theta_i^2 \to 0}\mathcal{F}_1(\theta_g^2,\theta_i^2)=
     \lim_{\theta_g^2/\theta_i^2 \to 0}\mathcal{F}_2(\theta_g^2,\theta_i^2)=0,
\end{align}
so that these contributions disappear asymptotically. 
We note that this result is not related to the clustering conditions but rather to the requirement that one of the two gluons ($k_1$) be within the groomed jet radius, i.e.\ $0<\theta_1^2<\theta_g^2$. Changing the integration variable to $\bar\theta_1^2$, we have $\theta_i^2<\bar \theta_1^2<\theta_g^2+\theta_i^2$. Thus, we have to evaluate the integral of a (regular or integrable) function over a domain that has a vanishing measure as $\theta_g^2 \to 0$.

As for the massless case, we do not perform the full NLL resummation of these effects, but we limit ourselves to exponentiate the $\mathcal{O}(\as^2)$ with running coupling corrections, as done in Eq.~(\ref{eq:clustering-running-massless}):
\begin{align}\label{eq:clustering-running-massive}
    S_i(\theta_g^2,\theta_i^2,\xi_b,\xi_c)= \exp &\Bigg[\frac{\cf^2 \mathcal{F}_1(\theta_g^2,\theta_i^2)  + \cf \ca \mathcal{F}_2(\theta_g^2,\theta_i^2)}{\pi^2}\nonumber\\ &\quad \quad \times\left(\int_{z_c \theta_g^\beta}^1 \frac{\de z}{z}\, \as(z^2 p_t^2 R_0^2)\right)^2\Bigg],
\end{align}

In summary, the resummed $\theta_g$ distribution for a heavy-flavour jet reads
\begin{equation}\label{eq:massiveNLL}
    \frac{1}{\sigma_0}\frac{\de \sigma_i}{\de \theta_g}=\frac{\de}{\de \theta_g} \left[ S_i(\theta_g^2,\theta_i^2,\xi_b,\xi_c)\, e^{- R_i(\theta_g^2,\theta_i^2,\xi_b,\xi_c)}\right],
\end{equation}
for $i=b,c$. The Sudakov exponent $R_i$ is given in Eq.~(\ref{eq:massive-radiator-final}) and the clustering contribution $S_i$ in Eq.~(\ref{eq:clustering-running-massive}). 
Note that our NLL distribution is no longer normalised, as a consequence of the fact that, in the massive case, the resummed cumulative behaves as a constant for $\theta_g\to 0$, while it vanishes in the massless case, see Eq.~(\ref{eq:masslessNLL}).
Indeed, while in the massless case, Soft Drop with $\beta \ge 0$ behaves perturbatively as a groomer, i.e. within resummed perturbation theory, it always returns a jet with $\theta_g>0$, the quark mass provides an effective cutoff so that, there is a non-vanishing probability, given by 
\begin{equation}
S_i(0,\theta_i^2,\xi_b,\xi_c)\, e^{-R_i(0,\theta_i^2,\xi_b,\xi_c)}\sim 
e^{-\frac{\as \cf}{\pi\beta}\left(\log^2(\zcut\theta_i^\beta)-\log^2(\zcut)\right)},
\end{equation}
to find Soft Drop jets with $\theta_g=0$, even if we expect fixed-order corrections, as well as non-perturbative effects, to smear this effect out.
Thus, we could either use Soft Drop in grooming mode and supplement Eq.~(\ref{eq:massiveNLL}) with an endpoint, $\delta(\theta_g)$, contribution that ensures normalisation or consider the algorithm in tagging mode and discard jets that return $\theta_g=0$. Henceforth, we choose this second option.

Finally, we note that, in the calculations of the resummed exponent presented so far, we have always included virtual corrections, as well as the contributions from jets that fail Soft Drop. In tagging mode, one should discard those and, in principle, repeat the calculations. However, we observe that in resummed perturbation theory
\begin{equation}
\Sigma(\theta_g)\Big|_\text{tagging}=\Sigma(\theta_g)\Big|_\text{grooming}-\Sigma(0)\Big|_\text{grooming}.
\end{equation}
Thus, in practice, we can just use the results obtained so far, provided that we remove the $\theta_g=0$ contribution. 
Numerical results and their comparison to Monte Carlo simulations will be presented in the next section.

\subsection{Numerical results and comparison to Monte Carlo simulations}\label{sec:theta_g_MC}
We now provide numerical results for the $\theta_g$ distribution according to Eq.~(\ref{eq:massiveNLL}), and compare them to Monte Carlo (MC) simulations. To this purpose, we generate events  using \herwig, version 7.2.2 \cite{Bellm:2019zci}. To test the all-order behaviour of the observables we are interested in, we consider LO matrix elements for the hard process, dressed with the parton shower.~\footnote{In this paper we concentrate on an angular-ordered parton shower, see~\cite{Hoang:2018zrp} for detailed comparisons with analytic resummation. We have also tested our analytic predictions against the \herwig dipole shower \cite{Platzer:2011bc} and the \pythia \cite{Bierlich:2022pfr} one and we did not find significant differences. It would be interesting, in the future, to also compare to state-of-the-art simulations that also include matching to fixed order.} Hadronisation effects are included using the \herwig cluster hadronisation model, when explicitly declared, and the CT14 \cite{Dulat:2015mca} set of PDFs has been used throughout the paper. 

We simulate the inclusive production of a pair of oppositely charged muons in association with a jet, in proton-proton collisions at $13$ TeV centre-of-mass energy. The muon pair is required to have an invariant mass between 70 and 110 GeV. Jets are clustered using the anti-$k_t$ algorithm \cite{Cacciari:2008gp} with $R_0=0.4$ and then ordered in transverse momentum. The hardest jet containing a $b$ ($c$) quark, or a $B$ ($D$) hadron, is considered. To match the heavy quark/hadron with a jet we look at the closest (with respect to the jet axis) flavoured particle, starting from the highest in transverse momentum, with $p_t > 5$ GeV. Properties of the flavoured particle and the jet are extracted and analysed using \rivet \cite{Buckley:2010ar} and \fastjet \cite{Cacciari:2011ma}. The fiducial phase-space for the muons is defined by the following cuts: $p_{t,\mu} > 26$ GeV, $|\eta_{\mu}|<2.4$, while the jets are selected in the region defined by $|\eta_J|<2.4$. We consider three different transverse-momentum regions, i.e. $p_{t} \ge 50, 150$ and $300$ GeV. However, only $p_{t}\ge150$ GeV is shown in the main text, and the other cases can be found in \ref{app:theta_g-other-pt}.

\begin{figure*} 
\begin{center}
\includegraphics[width=0.49\textwidth]{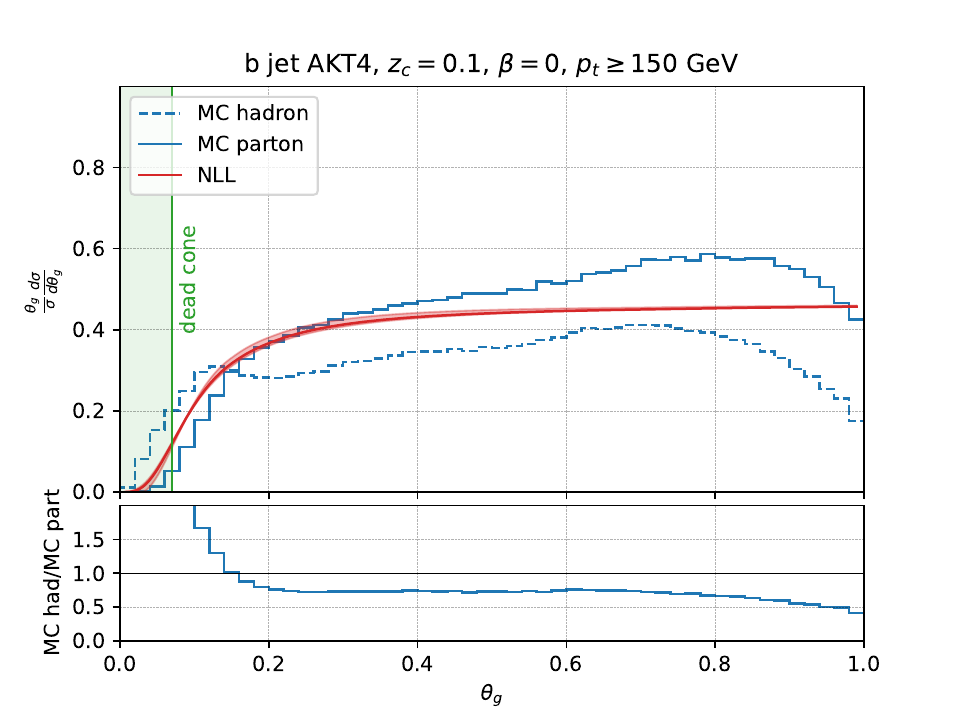}
\includegraphics[width=0.49\textwidth]{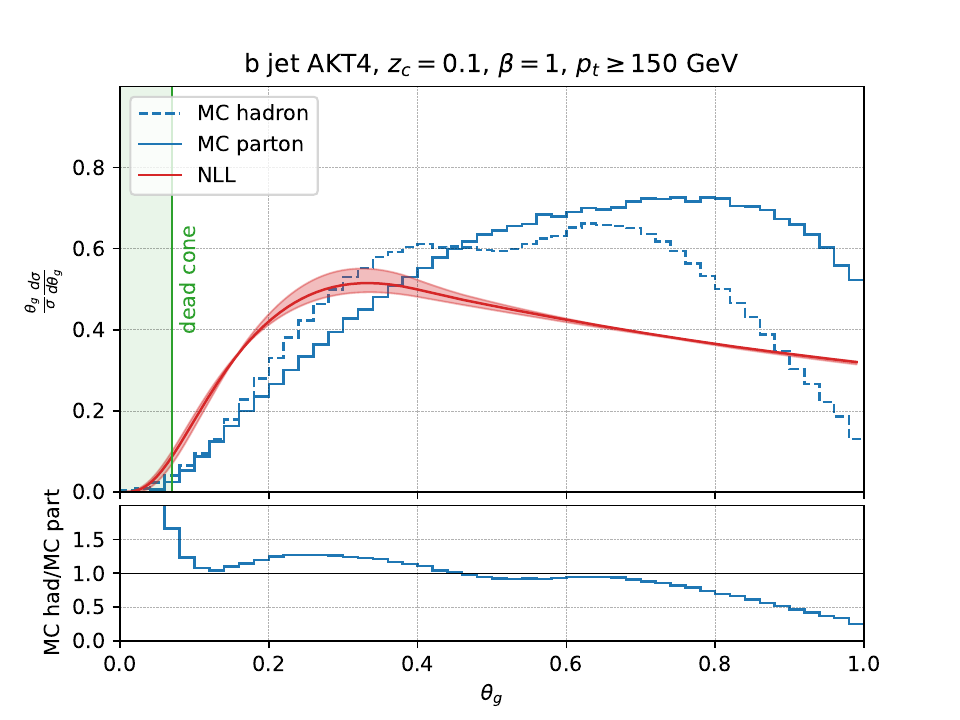}
\includegraphics[width=0.49\textwidth]{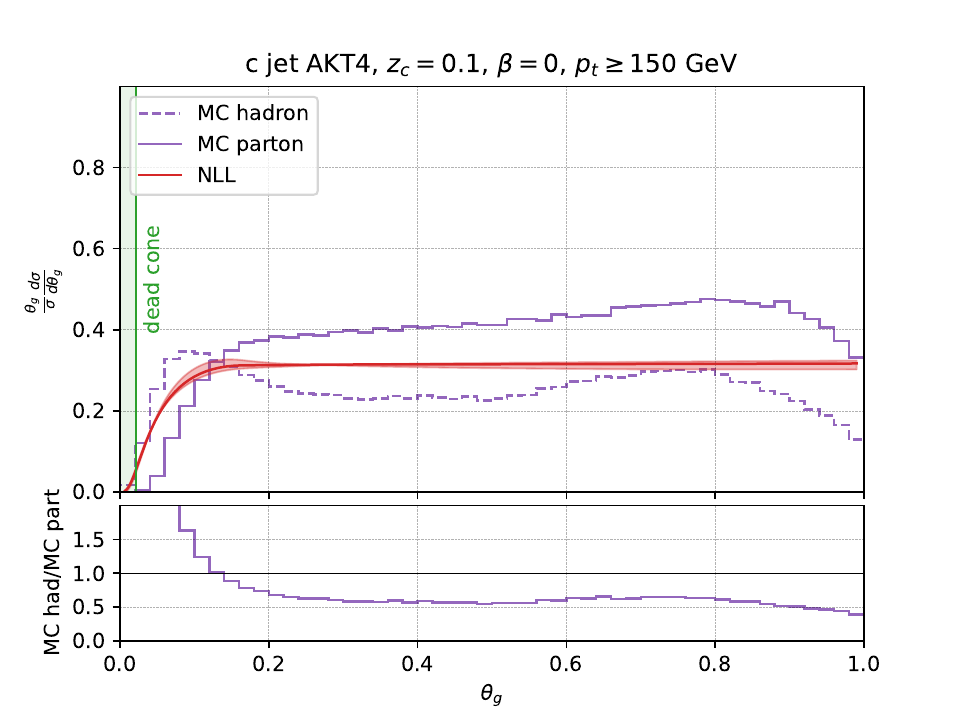}
\includegraphics[width=0.49\textwidth]{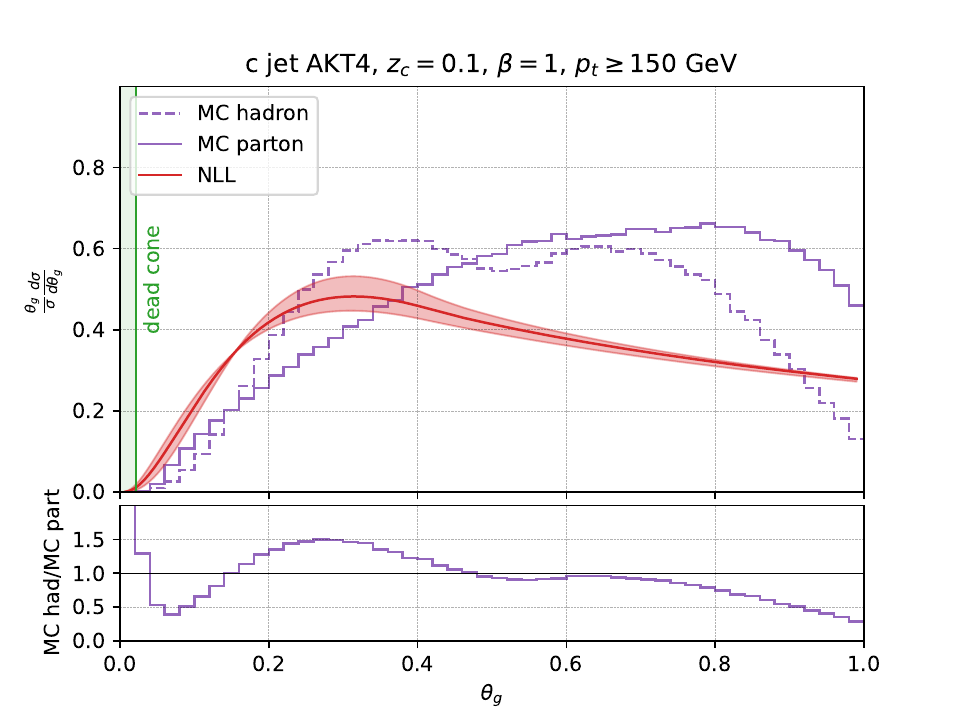}
\includegraphics[width=0.49\textwidth]{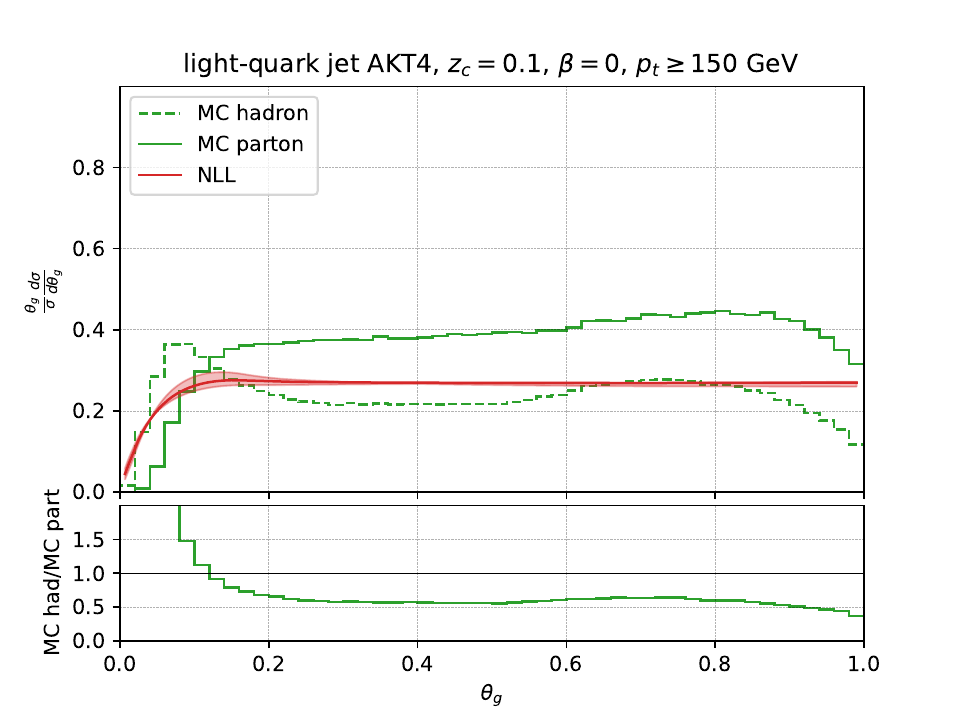}
\includegraphics[width=0.49\textwidth]{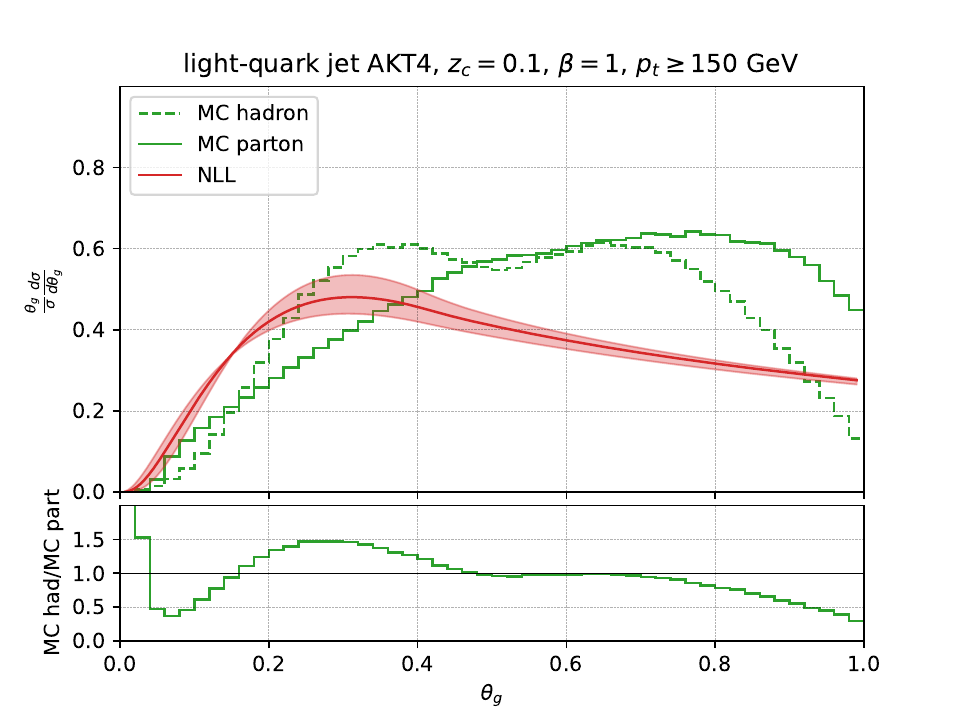}
\end{center}
\caption{The groomed jet radius ($\theta_g$) distribution for $b$-jets (top), $c$-jets (middle) and light-quark jets (bottom). In all plots, we show our NLL resummation, as well as the results obtained with the Monte Carlo event generator \herwig, both at parton and hadron level. Jets are selected with the anti-$k_t$ algorithm with $R_0=0.4$, with $p_t\ge 150$~GeV, and groomed with Soft Drop with $z_c=0.1$, and two values of the angular exponent: $\beta=0$ (on the left) and $\beta=1$ (on the right). The uncertainty bands for the analytic predictions are obtained by varying the resummation scale by a factor of two above and below the hard scale $p_t R_0$, i.e. $\mu_R \in \left[\frac{p_t R_0}{2}, 2 p_t R_0\right]$.}
\label{fig:thetag}
\end{figure*}

The $\theta_g$ distribution is shown in Fig.~\ref{fig:thetag}. Each plot includes parton level, i.e. with parton shower effect only, and hadron level, i.e. with hadronisation and the Underlying Event (UE) included. Together with the MC prediction, we show the corresponding NLL result of Eq.~(\ref{eq:massiveNLL}). The latter also exhibits a theoretical uncertainty given by the variation of the renormalisation scale, for which the central value is set at the hard scale $p_t R_0$, by a factor of two, as customary. Plots in different rows correspond to results for the identified leading jet in $Z+b$  ($B$), $Z+c$ ($D$) and $Z\,+$~light quark/hadron production, respectively. We always consider Soft Drop with $\zcut=0.1$, and $\beta=0$ and $1$, for left and right plots, respectively. For $b$ and $c$ jets, we also indicate, in green, the expected dead-cone region $\theta_g< \theta_i=\frac{m_i}{p_t R_0}$. All curves are normalised to have unit area. 
For $\beta=0$, i.e. when Soft Drop is more aggressive, we find good agreement between the NLL prediction and the MC. They are fairly different, instead, for $\beta=1$ at high $\theta_g$. However, in this region, we expect fixed-order corrections, not considered here, to be important. 

\begin{figure*} 
\begin{center}
\includegraphics[width=0.49\textwidth]{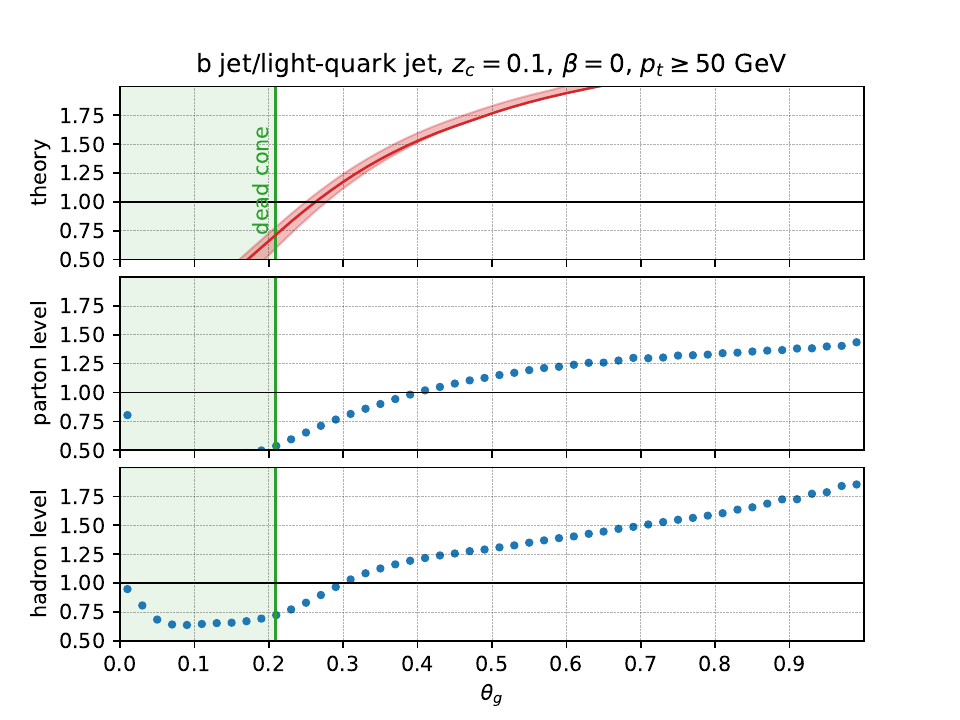}
\includegraphics[width=0.49\textwidth]{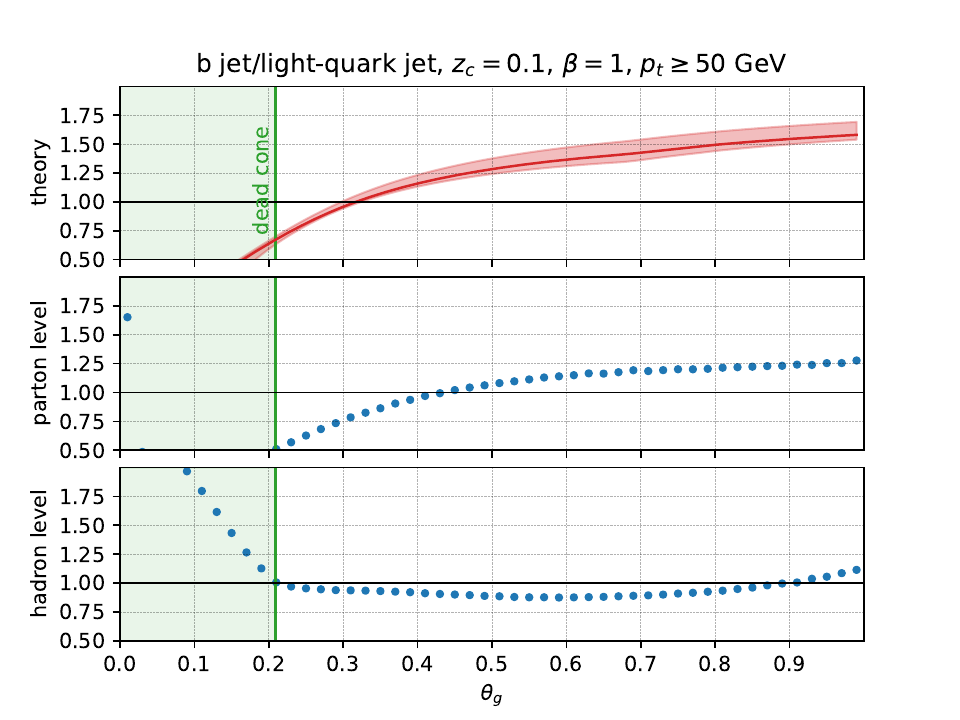}
\includegraphics[width=0.49\textwidth]{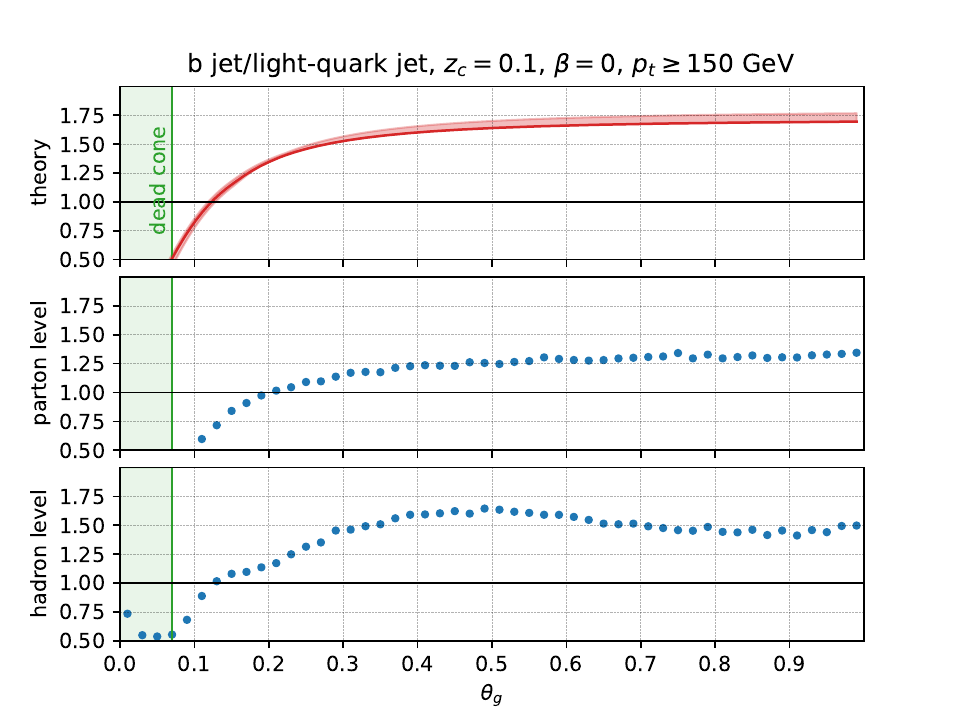}
\includegraphics[width=0.49\textwidth]{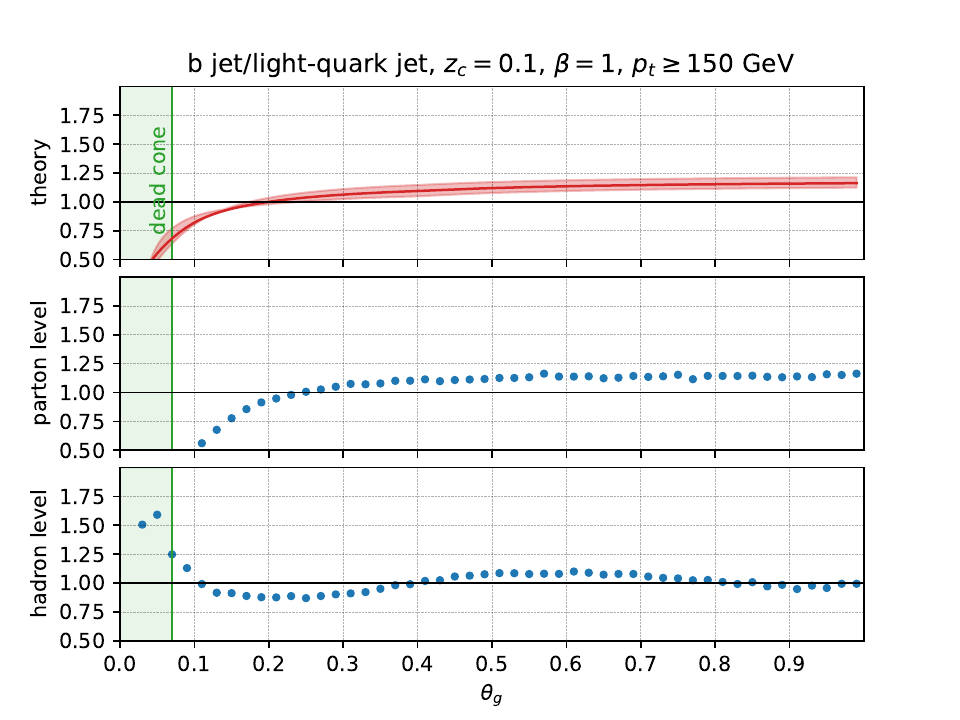}
\includegraphics[width=0.49\textwidth]{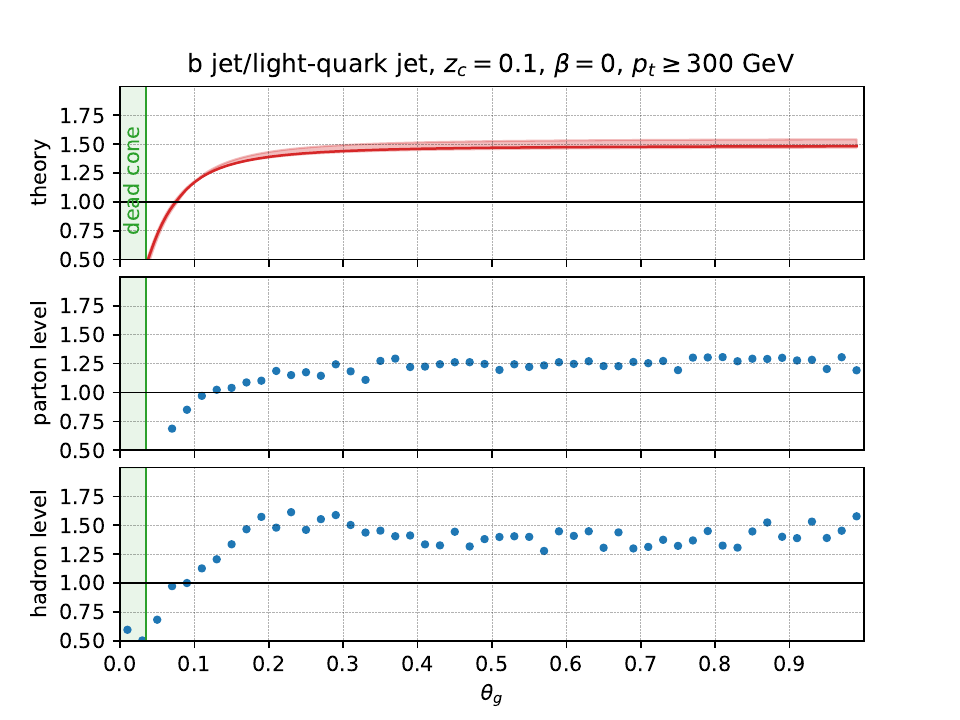}
\includegraphics[width=0.49\textwidth]{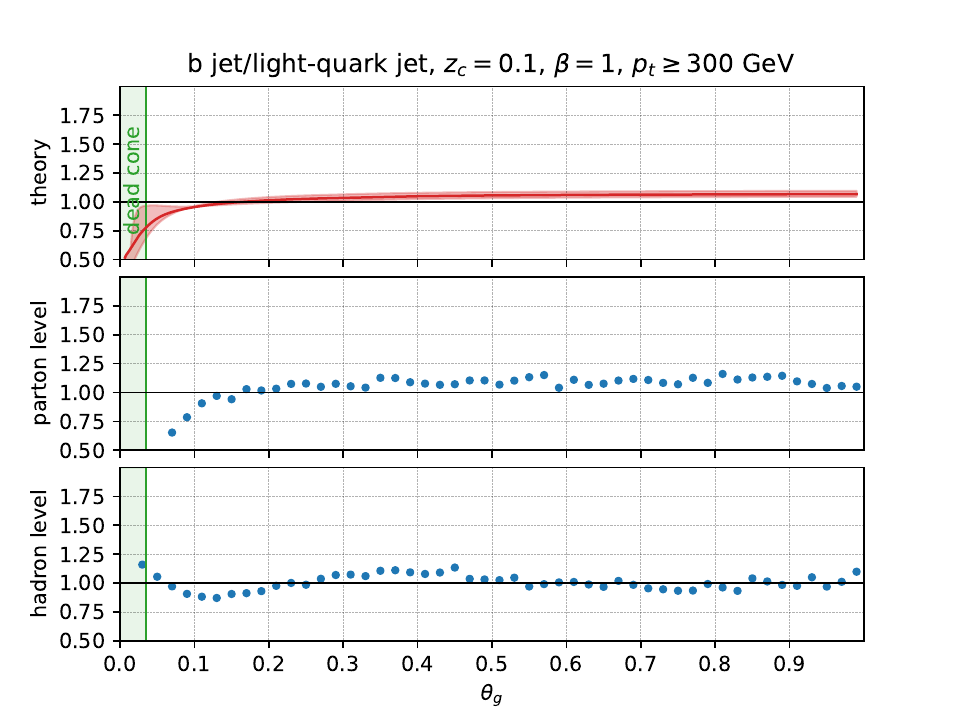}
\end{center}
\caption{Ratios of groomed jet radius ($\theta_g$) distribution of $b$ jets to light-quark jets for three values of the jet transverse momentum: $50$~GeV (top), $150$~GeV (middle) and $300$~GeV (bottom). The plots on the left are for $\beta=0$, and the ones on the right are for $\beta=1$. In each plot, we show our NLL resummation and the results obtained with \herwig, both a parton-level and hadron-level.
}
\label{fig:thetag-ratio}
\end{figure*}

In order to highlight possible dead-cone effects, it is useful to consider the ratio between the heavy-flavour jet $\theta_g$ distribution and the corresponding one for light-quark jets. We do this in Fig.~\ref{fig:thetag-ratio}. We concentrate on $b$-jets, but we show the results for three different transverse momentum cuts, namely 50, 150 and 300 GeV (from top to bottom). We show results for both $\beta=0$ (left) and $\beta=1$ (right). In every plot, we show the ratio obtained with our NLL resummed prediction and with the simulation performed with \herwig, both at parton-level and hadron-level. 

As expected,  because $\theta_i=\frac{m_i}{p_t R_0}$, the dead cone is more visible at lower values of $p_t$. Interestingly, deviations between the $b$ and the light quark distribution start at angular scales bigger than $\theta_i$. However, as $p_t$ is increased, this transition is pushed to a region that is likely beyond experimental resolution.
We also note that mass effects are more pronounced for the $\beta=0$ case than $\beta=1$. Considering that we have already noted that the former is under better theoretical control than the latter, the Soft Drop jets with $\beta=0$ appear to be an interesting choice to study the dead cone.

\section{The $z_g$ distribution}\label{sec:z_g}
We now discuss the second variable that characterises Soft Drop jets, namely the momentum fraction $z_g$. In this discussion, we want to provide a simple theoretical prediction for $z_g$ for jets initiated by massive quarks.
We also compare, both analytically and in Monte Carlo simulation, the variable $z_g$ to a widely-used fragmentation variable that measures the transverse momentum fraction of the $B(D)$ hadrons (or $b(c)$ quark at parton level) with respect to the jet $p_t$.

\subsection{Recap of the massless calculation}\label{sec:z_g_massless}

We start by briefly reviewing the calculation of the $z_g$ distribution for light jets.
The value of $z_g$ is fixed by the first de-clustering of the
jet that passes the \SD condition. Because we are completely inclusive
over the splitting angle, we must integrate over all possible values of $\theta_g$, including configurations where the two
emissions become collinear, for which the integral diverges. 
If $\beta \ge 0$, collinear splittings always pass the
\SD condition and these divergent configurations are not
cancelled by the corresponding virtual corrections, for which $z_g$ is
undefined, and heralds the fact that the observable is not IRC safe.

However, $z_g$ belongs to a wider class, i.e.\ Sudakov safe observables~\cite{Larkoski:2013paa,Larkoski:2014wba,Larkoski:2015lea,Komiske:2020qhg}, that despite being IRC unsafe, can be computed in perturbation theory, provided that we use resummation.
For this purpose, we need to introduce a safe companion observable. The \SD procedure itself suggests using the groomed angle $\theta_g$, which we have discussed in the previous section. 
Therefore, we imagine to measure a value of $z_g$, given a finite angular separation $\theta_g$. 
Using the language of conditional probabilities, we have~\cite{Larkoski:2015lea}:
\begin{equation}\label{eq:zg-cond-prob}
\frac{1}{\sigma_0}\frac{\de \sigma}{\de z_g} = \int_0^1 d \theta_g \, p(\theta_g) \, p(z_g|\theta_g),
\end{equation}
where $p(\theta_g)=\frac{1}{\sigma_0}\frac{\de \sigma}{\de \theta_g}$ and $p(z_g|\theta_g)=\frac{p(z_g,\theta_g)}{p(\theta_g)}$ is the conditional probability for measuring $z_g$ given a value of $\theta_g$.
If $\beta<0$, $z_g$ is IRC safe and the integral in Eq.~(\ref{eq:zg-cond-prob}) can be computed order by order in $\as$. This is no longer true when $\beta\ge0$, which is the standard configuration in which the Soft Drop algorithm is used and, therefore, our case of interest.  In this situation, the integral~(\ref{eq:zg-cond-prob}) diverges order by order in the strong coupling because of the $1/\theta_g$ behaviour of the integrand. However, if we take $p(\theta_g)$ to be the resummed distribution, i.e.\ the derivative of Eq.~(\ref{eq:masslessNLL}), then the Sudakov form factor regulates the $\theta_g=0$ singularity, providing us with a finite answer for Eq.~(\ref{eq:zg-cond-prob}).

Because in this work we are not concerned with logarithms of $z_g$ and $z_c$, we evaluate the conditional probability at fixed-order, focusing on the case of a quark-initiated jet:~\footnote{
We note that the formalism we have just presented does not describe the situation in which we have a $q \to q g$ splitting, with the gluon harder than the quark, i.e.\ when the minimum function of the Soft Drop condition~(\ref{eq:sdcrit}) returns $1-z$. 
This case can be accounted for by introducing a symmetrised version of the splitting function~\cite{Larkoski:2015lea,Larkoski:2017bvj,Tripathee:2017ybi}, which is integrated over $0<z<1/2$.
For consistency, one should then make this replacement also in the calculation of the resummed $\theta_g$ distribution. In what follows, in order to streamline our discussion, we keep the standard version of the splitting function. 
}
\begin{equation}\label{eq:cond-explicit}
p(z_g|\theta_g ) =  \frac{P_{gq}(z_g)\as(z_g^2 \theta_g^2 p_t^2 R_0^2)}{\int_{z_c\theta_g^\beta}^{1} \de z \, P_{gq}(z) \as(z^2 \, \theta_g^2 p_t^2 R_0^2)} \Theta(z_g-\zcut \theta_g^\beta).
\end{equation}
The Soft Drop condition~(\ref{eq:sdcrit}) dictates $z_g<\frac{1}{2}$ and, if $\beta=0$, $z_g>z_c$.

The integral in Eq.~(\ref{eq:zg-cond-prob}) with running coupling must be performed numerically. However, it is interesting to study its fixed-coupling and lowest-order approximation, see Eq.~(\ref{eq:radiator-fc-shift}). Thus, we consider
\begin{align}\label{eq:thetag-fc-simple}
    R^{(\text{f.c.})}(\theta_g^2)&= \frac{\as \cf}{\pi\beta}\Big(\log^2(\zcut\theta_g^\beta)-\log^2(\zcut)\Big)
    \nonumber\\ &=
     \frac{\as \cf}{\pi}\Big(\frac{\beta}{4} \log^2 \theta_g^2+  \log \theta_g^2 \, \log z_c\Big)
    , \nonumber\\  S^{(\text{f.c.})}(\theta_g^2)&=1.
\end{align}
For $\beta\ge 0$, we find~\cite{Larkoski:2015lea}
\begin{align}\label{eq:res_zf}
  &\frac{1}{\sigma_0}\frac{\de \sigma^{(\text{f.c.})}}{\de z_g}
  = P_{gq}(z_g)\frac{\as}{2\pi}
  \int_0^1\frac{\de \theta_g^2}{\theta_g^2}
  e^{-R(\theta_g^2)}
  \Theta(z_g-\zcut\theta_g^\beta)\nonumber\\
  &= \sqrt{\frac{\as }{4\beta \cf}} P_{gq}(z_g) e^{\frac{\as \cf}{\pi \beta}\log^2 z_c} 
  \Bigg[\text{erf}\left( \sqrt{\frac{\as \cf}{\pi \beta}} \log a \right)+1 \Bigg],
\end{align}
where $\text{erf}(x)$ is the error function and $a=\min[z_g,z_c]$. 
If $\beta>0$, the above result is non-analytic in $\as$:
\begin{equation}
\frac{1}{\sigma_0}\frac{\de \sigma^{(\text{f.c.})}}{\de z_g}  =
\frac{1}{2}\sqrt{\frac{\as }{\beta \cf }}\, P_{gq}(z_g)+{\cal O}\left(\alpha_s\right).
\end{equation}

Even more interesting is the $\beta=0$ case. 
At fixed coupling, the conditional probability, Eq.~(\ref{eq:cond-explicit}), becomes 
independent of $\theta_g$ and factors out of the 
integration to give
\begin{equation}
\label{eq:beta_zero_pzg}
\frac{1}{\sigma_0}\frac{\de \sigma^{(\text{f.c.})}}{\de z_g} = \frac{P_{gq}(z_g)}{\int_{z_c}^{1} \de z \, P_{gq}(z)}\Theta(z_g>\zcut)+{\cal O}\left(\alpha_s\right).
\end{equation}
It can be shown that the $\beta = 0$ case does have a valid
perturbative expansion in $\as$, despite being
$\as$-independent at lowest order.
Eq.~(\ref{eq:beta_zero_pzg}) shows that the $\beta=0$ $z_g$-distribution is essentially driven by the
QCD splitting function~\cite{Larkoski:2015lea}. 
This observation has initiated numerous theoretical~\cite{Mehtar-Tani:2016aco,KunnawalkamElayavalli:2017hxo,Chang:2017gkt,Milhano:2017nzm} and experimental \cite{ATLAS:2019mgf,ALargeIonColliderExperiment:2021mqf,ALICE:2019ykw,STAR:2021kjt,CMS:2017qlm,Kauder:2017cvz,Kauder:2017mhg} studies (see also~\cite{Tripathee:2017ybi,Larkoski:2017bvj}) that aim to use $z_g$ as a probe of QCD dynamics, both in $pp$ and heavy-ion collisions.

\subsection{The $z_g$ distribution for a heavy-flavour jet} \label{sec:z_g_massive}
We now move to the case of $b$ and $c$ jets and we start by considering the simple $\mathcal{O}(\as)$ calculation of the $z_g$ distribution, in the quasi-collinear limit:
\begin{align}\label{eq:zg-mass-FO-start}
\frac{1}{\sigma_0}\frac{\de \sigma_i^{(\text{f.o.})}}{\de z_g}&=
\frac{\as}{2\pi} \int_{0}^{1} \frac{\de \theta^2}{\theta^2+ \theta^2_i}\int_{z_c  \theta^\beta}^1 \de z \,P_{gi}\left(z, (z\theta p_t R_0)^2\right) \nonumber \\ & \times 
\delta(z-z_g) \nonumber\\
&=\frac{\as}{2\pi} \int_{\theta_i^2}^{1} \frac{\de \bar \theta^2}{\bar \theta^2 }  \,\mathcal{P}_{gi}\left(z_g, \bar \theta^2 \right)
\Theta(z_g-z_c \bar \theta^\beta)\nonumber\\
&+ \text{NNLL},
\end{align}
where, as before, we have dropped the mass-dependent shift in the Soft Drop condition. Even before performing the integral, it is clear that the mass of the heavy quark, as one might have expected, regulates the collinear singularity and so the $z_g$ distribution is IRC safe, for every value of $\beta$. The computation of the integral is straightforward but the presence of the $\bar \theta^2$ dependent contribution in the splitting function complicates the result. However, to illustrate our point, it is sufficient to work at LL. Therefore, we approximate $\mathcal{P}_{gi}=2 \cf/z_g$, and we find
\begin{align}\label{eq:zg-mass-FO-end}
\frac{1}{\sigma_0}\frac{\de \sigma_i^{(\text{f.o.})}}{\de z_g}&=-
\frac{\as \cf}{\pi} \frac{1}{z_g}
\begin{cases}
       \log\theta_i^2 & z_g>z_c, \\ 
      \frac{2}{\beta}\log \frac{z_c\theta_i^\beta}{z_g}\, & z_c \theta_i^\beta<z_g< z_c,
    \end{cases}
\end{align}
Note that in the case $\beta=0$, we have $z_g>z_c$ and only the first term survives.
Albeit finite, this expression contains logarithms of $\theta_i^2$, which become large in the boosted regime.
The all-order resummation of logarithms of $z_g$ partially addresses this problem. Indeed, keeping our focus on the LL fixed-coupling approximation, we find the following result for the normalised cumulative distribution:
\begin{align}\label{eq:zg-LL-of-zg}
&\log \Sigma_i^{(\text{f.c.})}(z_g)
 =\frac{\as\cf}{\pi}\Bigg[-\log \theta_i^2\, \log z_g \,\Theta(z_g-z_c)\nonumber\\&+
   \left(\frac{1}{\beta}\log^2 \frac{z_g}{z_c}-\log \theta_i^2\, \log z_g\right) \Theta(z_c-z_g)\Theta(z_g-z_c \theta_i^\beta)  \nonumber\\
   &+\left(\frac{1}{\beta}\log^2 \theta_i^\beta- \log \theta_i^2\, \log z_c\theta_i^\beta \right) \Theta(z_c \theta_i^\beta-z_g)\Bigg].
\end{align} 
The $\beta=0$ case is rather simple
\begin{align}\label{eq:zg-LL-of-zg-beta0}
\frac{1}{\sigma_0}\frac{\de \sigma_i^{(\text{f.c.})}}{\de z_g}&=-\frac{\as\cf }{\pi}\frac{\log \theta_i^2}{z_g}e^{-\frac{\as\cf}{\pi}\log \theta_i^2\, \log z_g }, \quad z_g>z_c.
\end{align}
We note that these expressions indeed resum those logarithms of $\theta_i$ that are associated with logarithms of $z_g$. However, the $\theta_i \to 0$ limit is not smooth and we do not recover the massless result of Eq.~(\ref{eq:res_zf}). 
This is related to the non-commutativity of the soft and massless limits, discussed at length in~\cite{Gaggero:2022hmv,Ghira:2023bxr}.

Another way of resumming logarithms $\theta_i^2$ is to resort to the conditional probability procedure described above for the massless case. 
To illustrate the procedure, we repeat, for the massive case, the calculation that led to the fixed-coupling result in Eq.~(\ref{eq:res_zf}).
At LL, the splitting function can be approximated by its soft contribution. Moreover, at this accuracy $R_i(\theta_g^2,\theta_i^2,\xi_b,\xi_c)=R(\vi)$, with $R$ given by Eq.~(\ref{eq:thetag-fc-simple}) and $\vi \in [\theta_i^2,1]$. Therefore, for $\beta\ge 0$, we find
\begin{align}\label{eq:zg-mass-cond-fixed-coupling}
&\frac{1}{\sigma_0}\frac{\de \sigma_i^{(\text{f.c.})}}{\de z_g}=
 \frac{\as \cf}{\pi}\frac{1}{z_g}
  \int_{\theta_i^2}^1\frac{\de \bar \theta^2}{\bar \theta^2}
  e^{-R(\bar \theta^2)}
  \Theta\left(z_g-z_c {\bar \theta}^\beta\right)\nonumber\\
  &= \sqrt{\frac{\as \cf }{\beta }} \frac{1}{z_g} e^{\frac{\as \cf}{\pi \beta}\log^2 z_c} \\ &\times
  \Bigg[\text{erf}\left( \sqrt{\frac{\as \cf}{\pi \beta}} \log a \right) -\text{erf}\left( \sqrt{\frac{\as \cf}{\pi \beta}} \log \left(z_c \theta_i^\beta\right) \right) \Bigg], \nonumber
\end{align}
with $a=\min[z_g,z_c]$. In particular, for $\beta=0$, we find
\begin{align}\label{eq:zg-mass-cond-fixed-coupling-beta0}
&\frac{1}{\sigma_0}\frac{\de \sigma_i^{(\text{f.c.})}}{\de z_g}=
\frac{1}{z_g \log z_c}\left(e^{-\frac{\as \cf}{\pi}\log \theta_i^2\log z_c} -1\right), \quad z_g>z_c.
\end{align}
The first-order expansions of the above expressions agree with Eq.~(\ref{eq:zg-mass-FO-end}) and large logarithmic corrections in $\theta_i^2$ are resummed. Furthermore, if we take $\theta_i\to0$, we recover the massless distribution of Eq.~(\ref{eq:res_zf}).
We further note that Eq. (\ref{eq:zg-mass-cond-fixed-coupling-beta0}) is the same as its massless counterpart, but for the normalisation factor in brackets. This makes sense because in the LL fixed-coupling approximation, both distributions are driven by the most singular part of the splitting function, which is the same for $P_{gq}$, $P_{gb}$, and $P_{gc}$.

Because it has the correct $\theta_i \to 0$ limit, we decide to use the conditional-probability approach to compute the $z_g$ distribution of a heavy-flavour jet:
\begin{align}\label{eq:zg-massiveNLL}
    \frac{1}{\sigma_0}\frac{\de \sigma_i}{\de z_g}&=\frac{1}{\sigma_0} \int_{\theta_i^2}^1 \de \bar \theta^2
     \frac{\mathcal{P}_{gi}(z_g,\bar \theta^2)\as(z_g^2 \bar \theta^2 p_t^2 R_0^2)}{\int_{\zcut\bar \theta^\beta}^{1} d z \, \mathcal{P}_{gi}(z,\bar \theta^2) \as(z^2 \, \bar \theta^2 p_t^2 R_0^2)}\nonumber \\ & \times \Theta(z_g-\zcut \bar \theta^\beta) \;
    \frac{\de \sigma_i}{\de \theta_g^2}\Bigg|_{\theta_g^2=\bar \theta^2-\theta_i^2}, \quad i=b,c,
    \end{align}
where the resummed $\theta_g$ distribution is given in Eq.~(\ref{eq:zg-massiveNLL}). 
We can make some further simplifications. The derivative in Eq.~(\ref{eq:zg-massiveNLL}) gives a factor that simplifies, within our accuracy, the denominator of the conditional probability. Thus, we obtain
    \begin{align}\label{eq:zg-massiveNLL-final}
    \frac{1}{\sigma_0}\frac{\de \sigma_i}{\de z_g}
    &=\frac{1}{\sigma_0} \int_{\theta_i^2}^1 \frac{\de \bar \theta^2}{\bar \theta^2}
     \mathcal{P}_{gi}(z_g,\bar \theta^2)\as(z_g^2 \bar \theta^2 p_t^2 R_0^2)  \nonumber \\ &\times S_i(\bar \theta^2,\theta_i^2,\xi_b,\xi_c)\, e^{- R_i(\bar \theta^2,\theta_i^2,\xi_b,\xi_c)}
    \Theta(z_g-\zcut \bar \theta^\beta).
\end{align}
Numerical results and their comparison to Monte Carlo simulations will be presented in the next section.

We close this discussion by noting that Eq.~(\ref{eq:zg-massiveNLL}) does not systematically resum logarithms of $z_g$ and $z_c$. This is acceptable for our purposes because we are mostly interested in the $\beta=0$ case, for which $z_g>z_c =0.1$. However, it would be interesting to extend the full NLL resummation for $z_g$, performed for light jets in~\cite{Cal:2021fla} to the massive case. Because the calculation in~\cite{Cal:2021fla} is based on the resummation of the double differential ($\theta_g,z_g)$, it may overcome the difficulties about the $\theta_i \to 0$ limit of Eqs.~(\ref{eq:zg-LL-of-zg}) and~(\ref{eq:zg-LL-of-zg-beta0}), previously discussed.

\subsection{Numerical results and comparison to Monte Carlo simulations}\label{sec:x_g_MC}
\begin{figure*} 
\begin{center}
\includegraphics[width=0.49\textwidth]{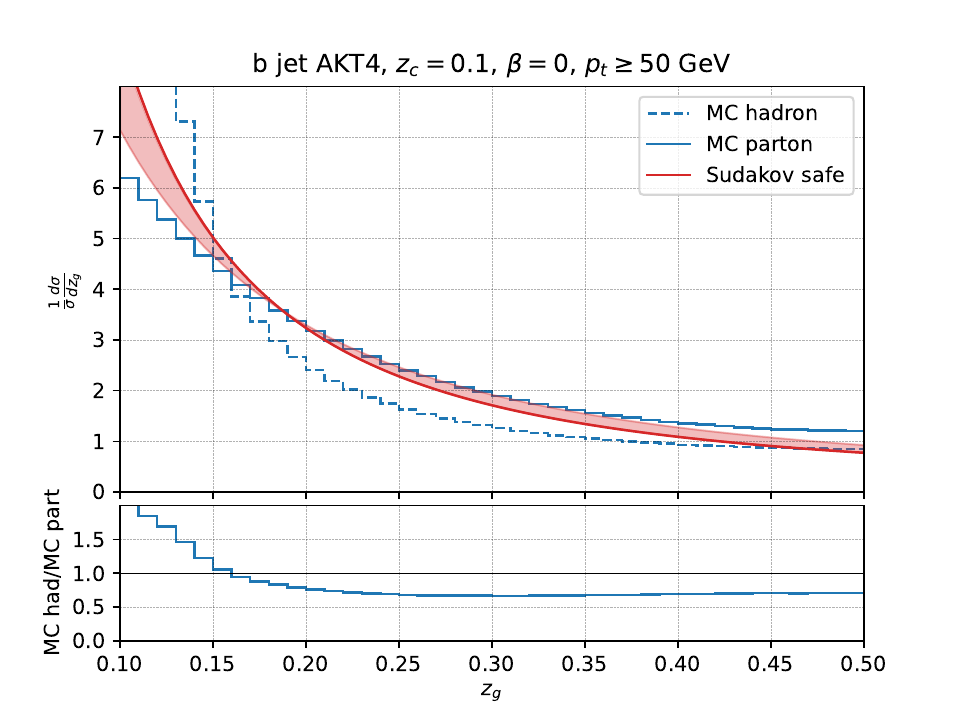}
\includegraphics[width=0.49\textwidth]{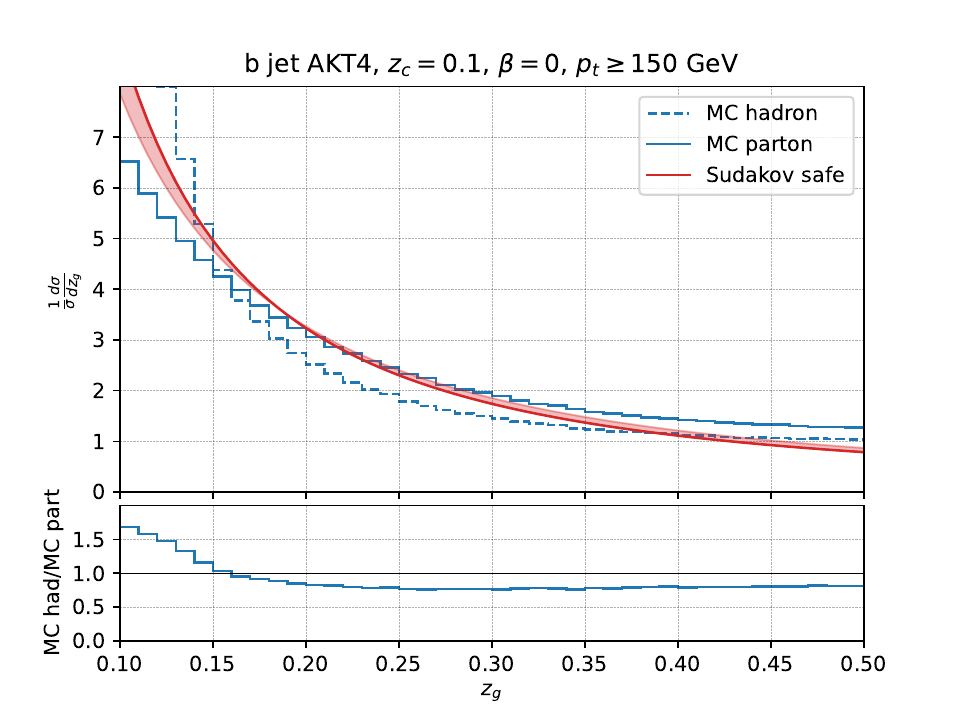}
\includegraphics[width=0.49\textwidth]{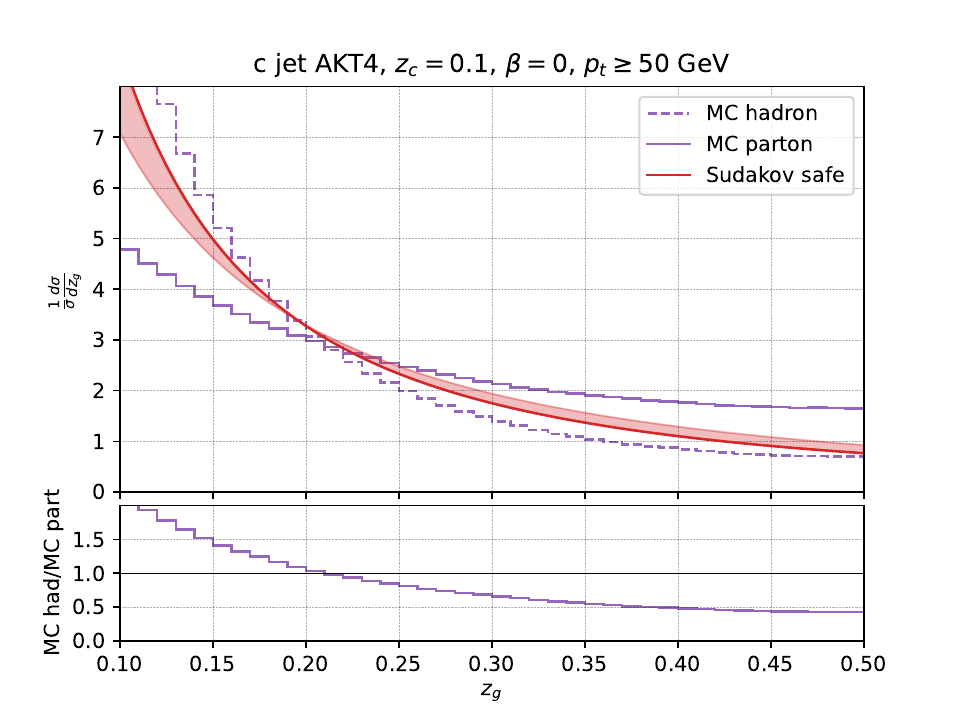}
\includegraphics[width=0.49\textwidth]{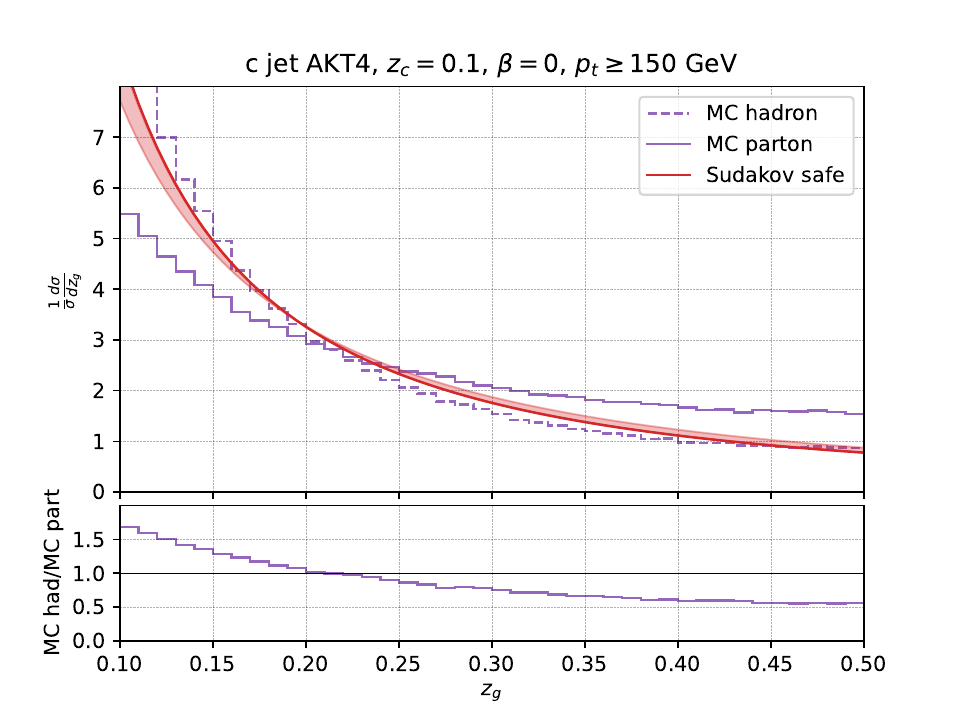}
\includegraphics[width=0.49\textwidth]{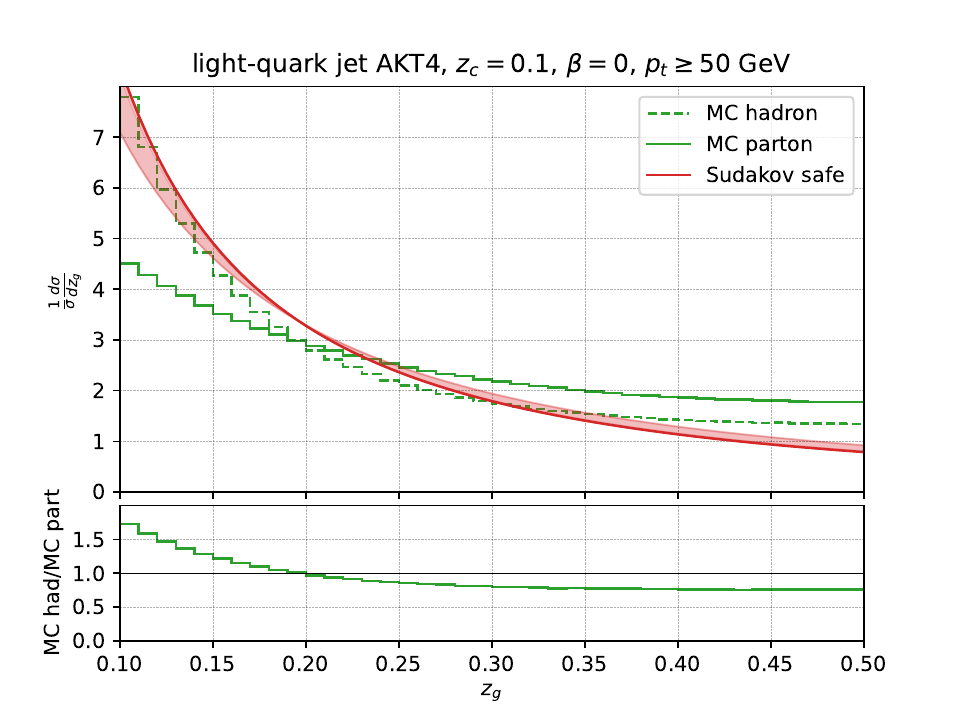}
\includegraphics[width=0.49\textwidth]{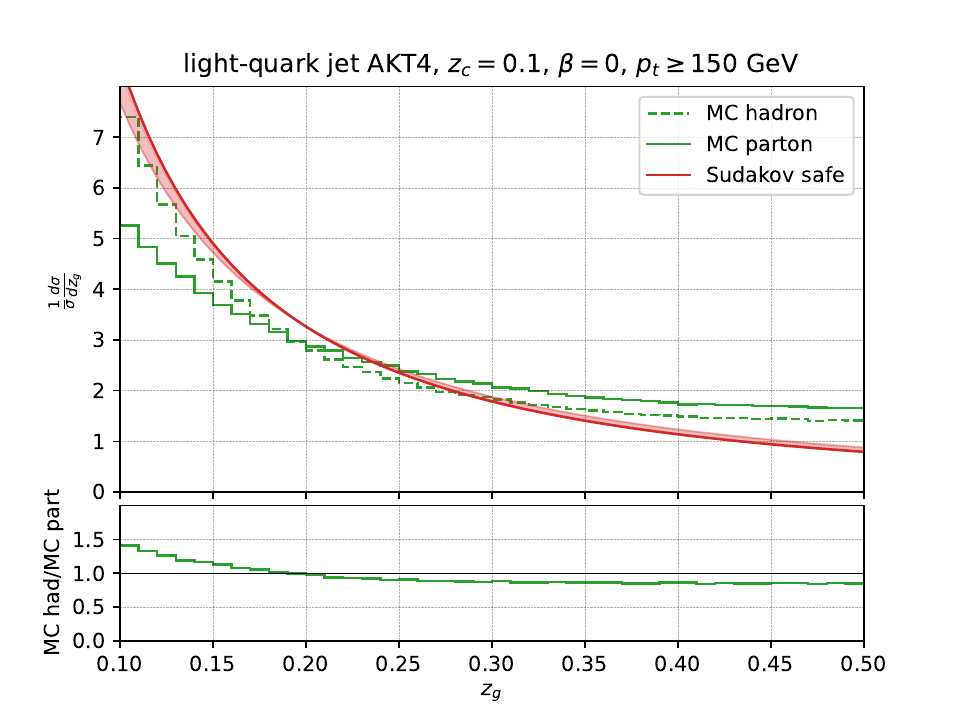}
\end{center}
\caption{The groomed momentum fraction ($z_g$) distribution for $b$-jets (top), $c$-jets (middle) and light-quark jets (bottom). In all plots, we show our resummed result, as well as the one obtained with the Monte Carlo event generator \herwig, both at parton and hadron level. Jets are selected with the anti-$k_t$ algorithm with $R_0=0.4$, and groomed with Soft Drop with $z_c=0.1$ and $\beta=0$. The plots on the left are for $p_t \ge 50$~GeV, while the ones on the right are for $p_t\ge 150$~GeV. The uncertainty bands for the analytic predictions are obtained by varying the resummation scale by a factor of two above and below the hard scale $p_t R_0$, i.e. $\mu_R \in \left[\frac{p_t R_0}{2}, 2 p_t R_0\right]$.}
\label{fig:zg}
\end{figure*}
In this section, we compare our resummed results for the $z_g$ distribution to the ones obtained with Monte Carlo event generators, both at parton- and hadron-level. We limit ourselves to the case $\beta=0$ and we simulate events using \herwig, with the same settings as the ones described in Sec.~\ref{sec:theta_g_MC}. 

In Fig.~\ref{fig:zg}, we compare resummed results for two different values of the jet transverse momentum cut, namely $p_t\ge50$ GeV, on the left, and $p_t\ge 150$ GeV, on the right. The plots at the top are for $b$-jets, the ones in the middle for $c$-jets, and the ones at the bottom for light-quark jets. All curves are normalised to have unit area. As already pointed out, the results all look very similar, because the shape of the distribution is mostly driven by the singular part of the splitting function for the emission of a gluon off a (massive) quark. However, we also note that differences between $b$,$c$, and light flavours are larger in the Monte Carlo results than in the analytic ones. This could be due to
quark masses influencing the kinematics, thus causing mass-dependent recoil effects. These are, to a certain extent, accounted for in the parton shower, but neglected in the analytic resummation.

Overall, we find good agreement between our calculation and the Monte Carlo results, although our results are strangely closer to the full simulation than to the parton-level one. We note that our predictions undershoot the Monte Carlo at large $z_g$. We have traced this back to the fact that our calculation does not take into account symmetrised splitting functions, as discussed in the footnote of Sec.~\ref{sec:z_g_massless}. Indeed by rescaling the $z_g$ distribution by the symmetrised splitting function, the tail of the analytic calculation moves closer to the Monte Carlo results. 

\subsection{Comparison to fragmentation functions}\label{sec:FF}

\begin{figure*} 
\begin{center}
\includegraphics[width=0.49\textwidth]{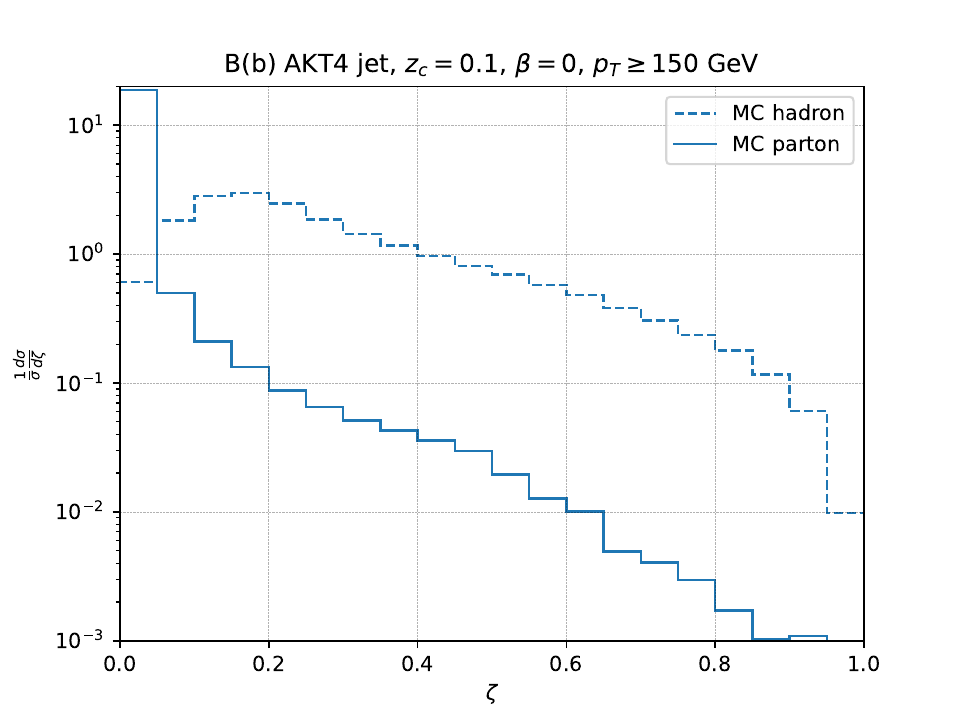}
\includegraphics[width=0.49\textwidth]{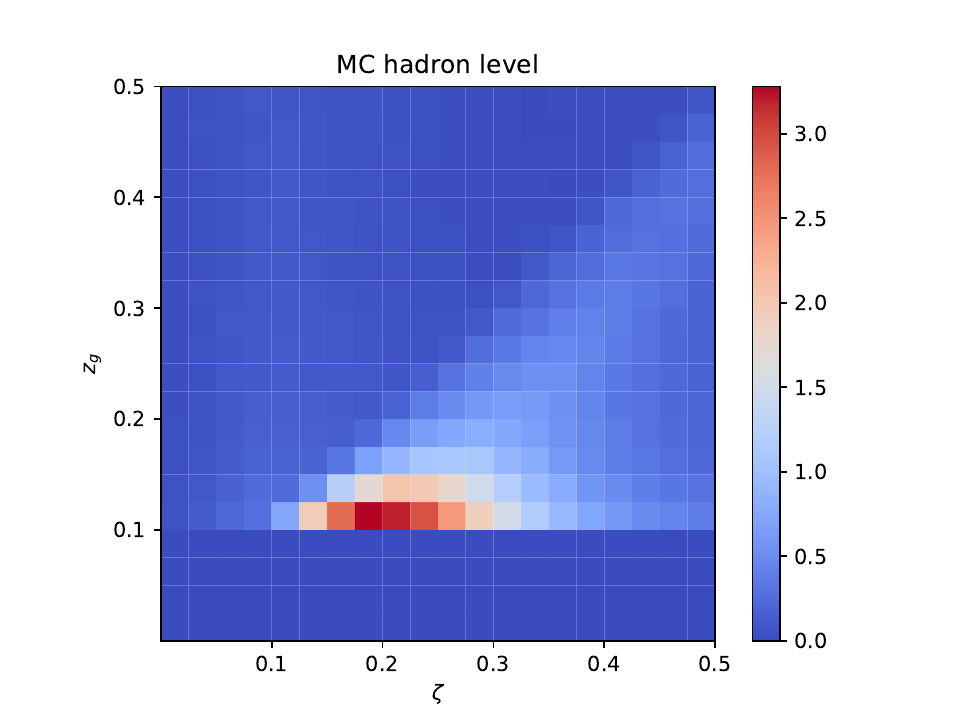}
\end{center}
\caption{Monte Carlo studies of the fragmentation function variable $x$, obtained with \herwig.
The plot on the left shows the $\zeta=1-x$ distribution for $b$-jets, at parton- and hadron-level. 
The plot on the right shows the correlation between $z_g$ (with $z_c=0.1$, and $\beta=0$) and $\zeta$, at hadron level.
In both cases, we have $p_t\ge 150$~GeV, and $R_0=0.4$.
}
\label{fig:zeta}
\end{figure*}

The study of heavy-flavour production at high energies is a multi-scale problem and, as already pointed out, logarithms of the ratio of the quark mass to the hard scale of the process can spoil the convergence of the perturbative expansion. 
Because these logarithmic corrections are related to collinear dynamics, heavy-flavour production cross-sections obey a factorisation theorem and they can be written as the convolution of process-dependent partonic (massless) coefficient functions with universal heavy-flavour fragmentation functions. Fragmentation functions obey DGLAP evolution equations with timelike splitting functions, which allow one to resum these large logarithmic corrections to all perturbative orders.

The initial condition for heavy quark fragmentation functions can be computed in perturbation theory, as originally pointed out in Ref.~\cite{Mele:1990cw,Mele:1990yq}, where the next-to-leading order (NLO) computation in QCD was presented.
The NNLO corrections were computed later in Refs.~\cite{Melnikov:2004bm,Mitov:2004du}. 
Furthermore, the initial condition of the evolution is affected by soft logarithms, that can be resummed to all orders too.~\cite{Cacciari:2001cw,Fickinger:2016rfd,Maltoni:2022bpy,Czakon:2022pyz,Ghira:2023bxr}.
Fragmentation functions for $b$ (or $c$) quarks are usually then supplemented with non-perturbative corrections before being compared to experimental data.
In the context of heavy-flavour production in $e^+ e^-$ collision, one of the most widely studied observables is perhaps the energy fraction $x$ of the heavy quark (or hadron) with respect to the energy of the incoming beam, in the centre-of-mass frame, see~\cite{Corcella:2022zna} (and references therein) for a recent review.

Collinear factorisation ensures that fragmentation functions are universal objects and so they can be used, in principle, to describe heavy flavours in hadronic collisions. However, a few changes in their definition are usually required from a practical point of view. For instance, although is possible to study directly the properties of identified $B$ or $D$ decays, the LHC experiments typically measure properties of the $B$ and $D$ hadrons within jets~\cite{ATLAS:2021agf,ATLAS:2022miz,LHCb:2017llq,ALICE:2019cbr}.
Theoretical studies of heavy quarks fragmenting in jets have been performed using effective field theories~\cite{Chien:2015ctp,Bain:2016clc,Bain:2017wvk,Kang:2017yde}.

In hadronic collisions, one usually measures projections of the hadron momentum with respect to the jet one (see e.g.~\cite{ATLAS:2021agf}) or the transverse momentum fraction of the heavy flavour with respect to the jet transverse momentum (see e.g.~\cite{ATLAS:2022miz}):
\begin{equation}
    x= \frac{p_{ti}}{p_t}, \quad i=b,c.
\end{equation}
Because $x=1$ at the Born level, if we want to study soft and collinear emissions, it is useful to define $\zeta=1-x$.

In what follows, we would like to study the correlation, if any, between the variable $\zeta$, which measures the departure from Born kinematics of the identified heavy flavour within a jet, to $z_g$ (with $\beta=0$), which probes the kinematics of the splitting that passes Soft Drop, in a heavy-flavour tagged jet.

Let us start with the $\mathcal{O}(\as)$ calculation, in the quasi-collinear limit. The result for $z_g$ can be immediately read off Eq.~(\ref{eq:zg-mass-FO-start}) by setting $\beta=0$:
\begin{align}\label{eq:zg-mass-FO-beta0}
\frac{1}{\sigma_0}\frac{\de \sigma_i^{(\text{f.o.})}}{\de z_g}&=
\frac{\as}{2\pi} \int_{\theta_i^2}^{1} \frac{\de \bar \theta^2}{\bar \theta^2 }  \,\mathcal{P}_{gi}\left(z_g, \bar \theta^2 \right)
\Theta(z_g-z_c).
\end{align}

The computation of the momentum fraction $\zeta$, in the same approximation, is rather straightforward. We consider the emission of a gluon with momentum fraction $z$ in the quasi-collinear limit. If the gluon is emitted within the jet, then $z=\zeta$, i.e. the $b(c)$ quark has momentum fraction $1-z=x$. If the gluon is emitted outside the jet, it does not contribute to the observable and $x$ remains equal to unity.~\footnote{This is different in $e^+e^-$ collision where the energy fraction is defined globally and gluon radiation always contributes.} We have
\begin{align}\label{eq:x-mass-FO-end}
\frac{1}{\sigma_0}\frac{\de \sigma_i^{(\text{f.o.})}}{\de \zeta}&=
\frac{\as}{2\pi} \int_{0}^{1} \frac{\de \theta^2}{\theta^2+ \theta^2_i}\int_0^1 \de z \,P_{gi}\left(z, (z\theta p_t R_0)^2\right) 
\delta(z-\zeta) \nonumber\\
&=\frac{\as}{2\pi} \int_{\theta_i^2}^{1} \frac{\de \bar \theta^2}{\bar \theta^2 }  \,\mathcal{P}_{gi}\left(\zeta, \bar \theta^2 \right),
\end{align}
where the second equation holds up to power corrections in the mass. 
Eq.~(\ref{eq:x-mass-FO-end}) coincides with Eq.~(\ref{eq:zg-mass-FO-beta0}), for $\zeta>z_c$, i.e.\ $x<1-z_c$. We conclude that $1-x$ and $z_g$ are fully correlated at this perturbative order. It is interesting to investigate whether this remains true beyond that. We study this problem by performing a Monte Carlo study. 

Before doing so, we study the $\zeta=1-x$ spectrum using the same \herwig simulation setup described for our previous analyses. The results are shown in Fig.~\ref{fig:zeta}, on the left, where we plot the $\zeta$ distribution at parton- and hadron-level, for $b$-jets with $p_t \ge 150$~GeV. 
We note that non-perturbative effects are large. This is interesting because, while fragmentation functions are under better perturbative control than Soft Drop observables, the latter ones seem to be  more robust against non-perturbative corrections.

Finally, the right-hand plot of Fig.~\ref{fig:zeta} shows the two dimensional distribution for $z_g$ (with $z_c=0.1$ and $\beta=0$) and $\zeta$, again for $p_t \ge 150$~GeV, at hadron-level. We note that the $\mathcal{O}(\as)$ correlation between these two observables is not maintained when higher-order corrections and non-perturbative effects are included. Therefore, we conclude that $z_g$ and $\zeta=1-x$ can offer different handles to study heavy-flavour dynamics at the LHC. 

\section{Conclusions and Outlook}\label{sec:conclusions}

In this work, we have considered heavy-flavour jets, namely $b$- and $c$-jets. In particular, we have studied heavy-flavour jets in conjunction with the Soft Drop grooming algorithm. 
Exploiting resummed perturbation theory, we have computed the NLL distribution for the groomed jet radius $\theta_g$, for both $b$- and $c$-jets. Our calculation accounts for the dead-cone effect due to the heavy-quark mass, as well as $m_b$ and $m_c$ thresholds in the running coupling, which is frozen at a non-perturbative scale $\Lambda=1$~GeV. We also discuss the role of clustering logarithms, first computed in~\cite{Kang:2019prh}.
The presence of many scales originates many cases and regions to be considered. We have listed all of them and shown numerical results for representative cases. 
We have compared our findings for the $\theta_g$ distribution with simulated data obtained with Monte Carlo event generators, assessing the role of non-perturbative corrections, such as hadronisation and the Underlying Event. We have found good agreement between resummed perturbation theory and Monte Carlo simulation, for the $\beta=0$ case, even for relatively low values of the hard scale, i.e. $p_t=50$~GeV and $R_0=0.4$. The agreement worsens if positive values of the angular exponent $\beta$ are considered.

We have also considered the momentum fraction $z_g$ of the first emission that passes Soft Drop. Our calculation generalises to the massive case the conditional-probability approach of~\cite{Larkoski:2015lea} and it allows for a resummation of mass effects. 
While our results hold for any $\beta \ge 0$, we have focussed our numerical investigations on the $\beta=0$ case. This choice is motivated by a few reasons. First, this is the value for which, at least to lowest order, one has a clear factorisation of the conditional probability expression, leading to a distribution proportional to the splitting function. For $\beta>0$, the dependence on the splitting function is smeared out. Second, it is for the $\beta=0$ case that one expects to find more similarities to the fragmentation variable $x$, and third, from the study of the $\theta_g$ distribution, we have found that the $\beta=0$ is under better theoretical control and more sensitive to dead cone effects.
We have compared our numerical results to Monte Carlo simulations. As for the massless case, the result is driven by the QCD splitting function and it is largely insensitive to non-perturbative effects. 

Finally, we have compared the $z_g$ distribution to the fragmentation variable $x$, which measures the ratio of the heavy quark (hadron) transverse momentum to the jet $p_t$. We performed the analytic calculation of both distributions at $\mathcal{O}(\as)$, showing that they lead to the same results for $z_g=1-x> z_c$. 
We have investigated the all-order behaviour of these observables using Monte Carlo simulations and discovered that the parton-shower significantly dilutes this correlation.
However, we have found that, in contrast to $z_g$, the $x$ receives sizeable non-perturbative corrections. 
In the future, it would be interesting to repeat these studies with the modified version of the declustering procedure~\cite{Cunqueiro:2018jbh}, whereby one follows the flavoured branch.

In this work, we have performed a detailed theoretical study of the effect of Soft Drop grooming on heavy-flavour jets. Before being able to compare to experimental data, such as the ones collected by the ALICE collaboration~\cite{ALICE:2022phr}, a few steps need to be taken. We are going to implement our calculation in the resummation plugin to \sherpa~\cite{Gerwick:2014gya, Reichelt:2021eru} in order to match our results to NLO theoretical predictions, with fiducial cuts, supplemented with non-perturbative corrections, as done for instance in~\cite{Marzani:2019evv,Baberuxki:2019ifp,Baron:2020xoi,Caletti:2021oor,Reichelt:2021svh,Knobbe:2023ehi}.
In this context, it would be also interesting to lift the small-$z_c$ limit and investigate, in the $\beta=0$ case, flavour-changing contributions that may induce radiation into the dead-cone region.
Finally, it would be important to improve the accuracy of the $z_g$ calculation by including the resummation of $z_g$ and $z_c$ logarithms, as done in~\cite{Cal:2021fla}.

\section*{Acknowledgments}
We thank Oleh Fedkevych, Silvia Ferrario Ravasio, Ezra Lesser, Davide Napoletano, Giovanni Ridolfi, Gregory Soyez, and Maria Ubiali for many inspiring discussions and comments on the manuscript.

We acknowledge support from the IPPP DIVA fellowship program and thank the physics department at the University of Cambridge for hospitality during the course of this work. 
AG and SM would like to thank the Erwin-Schr\"odinger International Institute for Mathematics and Physics at the University of Vienna for partial support during the Programme ``Quantum Field Theory at the Frontiers of the Strong Interactions", July 31 - September 1, 2023. 
SC would like to thank the University of Amsterdam and the Delta-ITP programme for support during the course of this work.
We also thank the Galileo Galilei Institute (GGI) for Theoretical Physics for the hospitality and the INFN for partial support during the completion of this work.

\appendix

\section{Running coupling integrals with quark-mass thresholds}\label{app:rc}
Throughout the paper, we have made use of the strong coupling in the Catani-Marchesini-Webber (CMW) scheme \cite{Catani:1990rr}:
\begin{align} \label{eq:as_CMW}
\as^{\text{CMW}}(k_t^2)=\as(k_t^2)\left(1+\as(k_t^2) \frac{K^{(n_f)}}{2\pi}\right),
\end{align}
where, in turn, $\as(k_t^2)$ is in the decoupling scheme:
\begin{align}\label{eq:flv-rc}
    \as(k_t^2)&=\as^{(5)}(k_t^2)\Theta(k_t^2-m_b^2) \nonumber\\&+
    \as^{(4)}(k_t^2)\Theta(k_t^2-m_c^2) \Theta(m_b^2-k_t^2) \nonumber\\&+
    \as^{(3)}(k_t^2)\Theta(k_t^2-\Lambda^2)\Theta(m_c^2-k_t^2)\nonumber\\
    & +\as^{(3)}(\Lambda^2)\Theta(\Lambda^2-k_t^2),
\end{align}
and
\begin{equation}\label{eq:K_CMW}
    K^{(n_f)}=\ca \left(\frac{67}{18}-\frac{\pi^2}{6}\right)-\frac{5}{9}n_f.
\end{equation}
In the above equation, $\as^{(n_f)}$ is the two-loop running coupling with $n_f$ active flavours. 
As a prescription to deal with the non-perturbative region, we have decided to freeze the coupling below  the scale $\Lambda \simeq 1$~GeV. 

In order to evaluate the integrals over the running coupling, we express each $\as^{(n_f)}(k_t^2)$ appearing in Eq.~(\ref{eq:flv-rc}) in terms of the value of the strong coupling at the hard scale $p_t R_0$, which we assume to be above the $b$-quark mass. Requiring continuity at the two quark-mass thresholds, at two loops, we find
\begin{align}
    \as^{(5)}(k_t^2)&= \frac{\as}{1+\nu_5}\left(1-\as \frac{\beta_1^{(5)}}{\beta_0^{(5)}}\frac{\log{(1+\nu_5)}}{1+\nu_5}\right),\\
    \as^{(4)}(k_t^2)&=\frac{\as}{1+\nu_4-\delta_{54}}\left(1-\as \frac{\beta_1^{(4)}}{\beta_0^{(4)}} \frac{\log{(1+\nu_4-\delta_{54})}}{1+\nu_4-\delta_{54}}\right)\nonumber \\
    &-\left(\frac{\beta_1^{(5)}}{\beta_0^{(5)}}-\frac{\beta_1^{(4)}}{\beta_0^{(4)}}\right)\log{(1-\lambda^{(5)}_{\xi_b})} \frac{\as^2}{(1+\nu_4-\delta_{54})^2},\\
    \as^{(3)}(k_t^2)&= \frac{\as}{1+\nu_3-\delta_{54}-\delta_{43}}\nonumber\\
    &\times \left(1-\as \frac{\beta_1^{(3)}}{\beta_0^{(3)}} \frac{\log{(1+\nu_3-\delta_{54}-\delta_{43})}}{1+\nu_3-\delta_{54}-\delta_{43}}\right)\nonumber\\
    &-\as^2\left(\frac{\beta_1^{(4)}}{\beta_0^{(4)}}-\frac{\beta_1^{(3)}}{\beta_0^{(3)}}\right) \frac{\log{\left(1-\delta_{54}-\lambda^{(4)}_{\xi_c}\right)}}{(1+\nu_3-\delta_{54}-\delta_{43})^2}\nonumber\\
    &-\as^2\left(\frac{\beta_1^{(5)}}{\beta_0^{(5)}}-\frac{\beta_1^{(4)}}{\beta_0^{(4)}}\right) \frac{\log{(1-\lambda^{(5)}_{\xi_b})}}{(1+\nu_3-\delta_{54}-\delta_{43})^2},
\end{align}
where $\as=\as^{(5)}(p_t^2 R_0^2)$ and we have introduced
\begin{align}
        \nu_{n_f}&=\as \beta^{(n_f)}_0 \log{\frac{k_t^2}{p_t^2 R_0^2}},\quad n_f=3,4,5,\\
        \lambda^{(n_f)}_{\xi_b}&= \as \beta^{(n_f)}_0 \log{\frac{1}{\xi_b}},\quad n_f=4,5,\\
         \lambda^{(n_f)}_{\xi_c}&= \as \beta^{(n_f)}_0 \log{\frac{1}{\xi_c}}, \quad n_f=3,4,\\
        \delta_{54}&= \lambda^{(5)}_{\xi_b}-\lambda^{(4)}_{\xi_b}, \quad \delta_{43}=\lambda^{(4)}_{\xi_c}-\lambda^{(3)}_{\xi_c}.
\end{align}
The two-loop coefficients of the QCD $\beta$-function are
\begin{align}
	\beta_0^{(n_f)}=\frac{11\ca-2 n_f}{12\pi},\quad \beta_1^{(n_f)}=\frac{17\ca^2-5\ca n_f-3\cf n_f}{24\pi^2},
\end{align}
with $\ca=3$ and $\cf=\frac{4}{3}$.

All the expressions we have to compute can be cast as a double integration over the emission's transverse momentum and its angle with respect to the hard quark, at fixed transverse momentum. Furthermore, it proves convenient to change integration variables to logarithmic ones, namely $\nu$ and the emission's rapidity. This way, the integral
over rapidity can always be written as $L_0 + c \nu$ with $L_0$ and $c$
independent of $\nu$.
Therefore, all these expressions have the form:
\begin{align} \label{eq:LL and part of NLL} 
I^{(n_f)}\left(\lambda_a,\lambda_b,L_0,c\right)&=\frac{\cf}{\left(\as \beta^{(n_f)}_0\right)^2} \int^{-\lambda_a}_{-\lambda_b} \de \nu \frac{\as^{\text{CMW}}(\bar k_t^2)}{2\pi} \nonumber \\ &\times \left(L_0+ c \nu\right),\qquad \bar k_t= p_t R_0 \exp\frac{\nu}{\as\beta^{(n_f)}_0 }.
\end{align}
The limits of integration, as well as $L_0$, are linear combinations of $\lambda^{(n_f)}=\as \beta^{(n_f)}_0 \log{\frac{1}{\vi}}$, $\lambda^{(n_f)}_{\text{cut}}=\as \beta^{(n_f)}_0 \log{\frac{1}{\zcut}}$, $\lambda^{(\text{NP})}=\as \beta_0^{(3)} \log{\frac{\left(p_t R_0\right)^2}{\Lambda^2}}$ and $\lambda^{(n_f)}_{\xi_i}$, defined above. In contrast, $c$ can only assume the values: $\left \{\frac{1}{1+\beta},-\frac{\beta}{1+\beta},-1,0\right\}$.

As long as $k_t^2>\Lambda^2$, the integral in Eq.~(\ref{eq:LL and part of NLL}) can be written as:
\begin{equation}
\begin{split} \label{eq: generating function}
    I^{(n_f)}(\lambda_a,&\lambda_b,L_0,c)= \frac{\cf}{2\pi\beta^{(n_f) 2}_0} \bigg(\frac{1}{\as}I_1(\lambda_a,\lambda_b,L_0,c)\\
    +&\frac{K^{(n_f)}_{\text{eff}}}{2\pi}I_2(\lambda_a,\lambda_b,L_0,c)-\frac{\beta_1^{(n_f)}}{\beta_0^{(n_f)}}I_3(\lambda_a,\lambda_b,L_0,c)\bigg),
    \end{split}
\end{equation}
with:
\begin{equation}
    \begin{split} \label{eq: results integral}
        I_1=& \left(L_0-c(1-x^{(n_f)})\right) \log{\left(\frac{1-x^{(n_f)}-\lambda_a}{1-x^{(n_f)}
        -\lambda_b}\right)}\\
        -&c \left(\lambda_a-\lambda_b\right),\\
        I_2=& (c(1-x^{(n_f)})-L_0)\frac{\lambda_a-\lambda_b}{(1-x^{(n_f)}-\lambda_a)(1-x^{(n_f)}-\lambda_b)}\\
        +&c\log{\left(\frac{1-x^{(n_f)}-\lambda_a}{1-x^{(n_f)}
        -\lambda_b}\right)}\\
        I_3=& (c(1-x^{(n_f)})-L_0)\\
        \times&\bigg(\frac{1+\log{\left(1-x^{(n_f)}-\lambda_a\right)}}{1-x^{(n_f)}-\lambda_a}-
        \frac{1+\log{\left(1-x^{(n_f)}-\lambda_b\right)}}{1-x^{(n_f)}-\lambda_b}\bigg)\\
        +&c \left(\frac{1}{2}\log^2{(1-x^{(n_f)}-\lambda_a)}-\frac{1}{2}\log^2{(1-x^{(n_f)}-\lambda_b)}\right),
    \end{split}
\end{equation}
and
 \begin{equation}
     x^{(n_f)}=
     \begin{cases}
         0 & \text{for}~n_f=5,\\
         \delta_{54} &\text{for}~n_f=4,\\
          \delta_{54}+\delta_{43} &\text{for}~n_f=3.
     \end{cases}
 \end{equation}
The coefficient $K^{(n_f)}_{\text{eff}}$ is defined as:
\begin{equation}
    K^{(n_f)}_{\text{eff}}=
    \begin{cases}
        K^{(5)} \quad \text{for}~n_f=5,\\
        K^{(4)}-2\pi\left(\frac{\beta_1^{(5)}}{\beta_0^{(5)}}-\frac{\beta_1^{(4)}}{\beta_0^{(4)}}\right)\log{(1-\lambda^{(5)}_{\xi_b})} \;\text{for}~n_f=4,\\
        K^{(3)}-2\pi\left(\frac{\beta_1^{(4)}}{\beta_0^{(4)}}-\frac{\beta_1^{(3)}}{\beta_0^{(3)}}\right)\log{\left(1-\delta_{54}-\lambda^{(4)}_{\xi_c}\right)}\\
        -2\pi\left(\frac{\beta_1^{(5)}}{\beta_0^{(5)}}-\frac{\beta_1^{(4)}}{\beta_0^{(4)}}\right)\log{(1-\lambda^{(5)}_{\xi_b})} \quad \text{for}~n_f=3.
    \end{cases}
\end{equation}
When $k_t$ becomes smaller than the non-perturbative scale $\Lambda$, we freeze the coupling at $\Lambda$, see Eq.~(\ref{eq:flv-rc}). In this case the result of Eq.~(\ref{eq:LL and part of NLL}) is straightforward, and reads:
\begin{equation}
    \begin{split}
\label{eq: non pert}
        I^{(\text{NP})}\left(\lambda_a,\lambda_b,L_0,c\right)&=\frac{\cf}{\left(\as \beta^{(n_f)}_0\right)^2}\frac{\as^{\text{CMW}}\left(\Lambda^2\right)}{2\pi} \\
        &
        \left(\frac{c}{2}(\lambda_a^2-\lambda_b^2)-L_0(\lambda_a-\lambda_b)\right)
    \end{split}
\end{equation}
In order to achieve NLL accuracy, two further contributions must be considered. These are obtained by integrating the less singular components of the splitting function. At this accuracy, we can evaluate the strong coupling at one loop. We must take into account the hard-collinear term of the splitting function:
\begin{equation}
    \begin{split}
    \label{eq:hard-collinear}
        B^{(n_f)}=& \frac{B_1}{2\pi \beta_0^{(n_f)} \as} \int^{-\lambda_a}_{-\lambda_b} \de \nu~ \as^{(n_f)}(\nu)=\\
        &\frac{B_1}{2\pi \beta_0^{(n_f)}}\log{\left(\frac{1-x^{(n_f)}-\lambda_a}{1-x^{(n_f)}
        -\lambda_b}\right)},
    \end{split}
\end{equation}
and mass-dependent one:
\begin{equation}
\begin{split}
\label{eq:dead-cone}
    H^{(n_f)}= \frac{H_1}{2\pi \beta_0^{(n_f)}\as}& \frac{\theta_i^2}{\vi} \int^{-\lambda_a}_{-\lambda_b} \de \nu~ \as^{(n_f)}(\nu)=\\
    &\frac{H_1 \theta_i^2}{2\pi \beta_0^{(n_f)} \vi}\log{\left(\frac{1-x^{(n_f)}-\lambda_a}{1-x^{(n_f)}
        -\lambda_b}\right)},
        \end{split}
\end{equation}
with $B_1= -\frac{3}{2}\cf,~ H_1=-\cf $. In the non-perturbative regions, both integrals in Eqs.~(\ref{eq:hard-collinear},\ref{eq:dead-cone}) reduce to the result in Eq.~(\ref{eq: non pert}) with the replacements $\as^{\text{CMW}}\to \as$, $L_0=1$, and $c=0$.

\section{Mapping out the different regions}\label{app:regions}
The integrals that are necessary in order to describe heavy-flavour jet distributions are all of the types described in the previous appendix. What makes the calculation cumbersome is the presence of multiple scales, which implies the appearance of many different regions and cases to be considered. 

The Soft Drop condition, the heavy-flavour thresholds and the non-perturbative scales determine 7 different values for the dimensionless transverse momentum $\kappa<1$ that appears on the vertical axis of the Lund plane in Fig.~\ref{fig:lundplane}:
\begin{align}
    \kappa^2&= z_c^2, \; \xi_i, \; z_c^2 \xi_i^{1+\beta}, \quad \text{with} \quad i=b,c,\Lambda,
\end{align}
Clearly, we have to take into account additional constraints that will reduce the number of possible cases much below $7!$. We have
\begin{equation}
    \begin{cases}
    &\xi_b> \xi_c> \xi_\Lambda, \nonumber \\
    & z_c^2 \xi_b^{1+\beta}> z_c^2 \xi_c^{1+\beta}> z_c^2 \xi_\Lambda^{1+\beta}, \nonumber\\
    & \xi_i > z_c^2 \xi_i^{1+\beta}.
\end{cases}
\end{equation}
Note that this implies that $z_c^2 \xi_\Lambda^{1+\beta}$ is always the smallest one.
We classify the different cases according to the value of $z_c$.
\begin{enumerate}
    \item[(a)] The case $z_c^2>\xi_b$ has the following subcases:
\begin{center}
\begin{tabular}{|c|c|c|c|c|}
\hline
 1 &$\xi_c$ & $\xi_\Lambda$ & $z_c^2\xi_b^{1+\beta}$ & $z_c^2\xi_c^{1+\beta}$\\ 
 \hline
  2 & $\xi_c$ & $z_c^2\xi_b^{1+\beta}$ & $\xi_\Lambda$ & $z_c^2\xi_c^{1+\beta}$\\  
  \hline
  3 &$\xi_c$ & $z_c^2\xi_b^{1+\beta}$ & $z_c^2\xi_c^{1+\beta}$ & $\xi_\Lambda$\\   
  \hline
  4 &$z_c^2\xi_b^{1+\beta}$ & $\xi_c$ & $\xi_\Lambda$ & $z_c^2\xi_c^{1+\beta}$\\
  \hline
  5 &$z_c^2\xi_b^{1+\beta}$ & $\xi_c$ & $z_c^2\xi_c^{1+\beta}$ &$\xi_\Lambda$\\ \hline
\end{tabular}
\end{center}
where it is understood that the values in the table are ordered from big to small.
\item[(b)] The case $\xi_b> z_c^2> \xi_c$ has the same subcases as (a).
\item[(c)] The case $\xi_c> z_c^2> \xi_\Lambda$ has 3 subcases:
\begin{center}
\begin{tabular}{|c|c|c|c|}
\hline
 1 & $\xi_\Lambda$ & $z_c^2\xi_b^{1+\beta}$ & $z_c^2\xi_c^{1+\beta}$\\ 
 \hline
  2 &  $z_c^2\xi_b^{1+\beta}$ & $\xi_\Lambda$ & $z_c^2\xi_c^{1+\beta}$\\  
  \hline
  3 & $z_c^2\xi_b^{1+\beta}$ & $z_c^2\xi_c^{1+\beta}$ & $\xi_\Lambda$\\   
  \hline
\end{tabular}
\end{center}
\item[(d)] Finally, for the last case, we can only have $ \xi_\Lambda>z_c^2>z_c^2 \xi_b^{1+\beta}>z_c^2 \xi_c^{1+\beta}$.
\end{enumerate}
Thus, we have to consider 14 cases. This is what actually happens for light jets and $c$-jets.
For $b$-jets, the situation is slightly simpler because the dead cone at $\theta_i^2=\xi_b$ implies that we are not sensitive to the scale $z_c^2 \xi_c^{1+\beta}$. Thus, we have only 3 distinct subcases for (a) and (b), 2 for (c) and still 1 for (d), totalling 9 cases for $b$-jets.

In this work, we have fixed the value of $z_c=0.1$ and considered only two distinct values for the angular exponent $\beta=0,1$. We have also worked with just one value for the anti-$k_t$ jet radius, $R_0=0.4$, and fixed the non-perturbative scale $\Lambda=1$ GeV. 
This results in only two different hierarchies to be considered. In fact, for $p_t= 150,300$ GeV we have:
 \begin{equation}
     \begin{split}
         \zcut^2>\xi_b> \xi_c> \xi_\Lambda> \zcut^2 \xi_b^{1+\beta}> \zcut^2 \xi_c^{1+\beta}>\zcut^2 \xi_\Lambda^{1+\beta},
     \end{split}
 \end{equation}
while, for $p_t= 50~ \text{GeV}$, we find
\begin{equation}
    \xi_b>\zcut^2>\xi_c> \xi_\Lambda> \zcut^2 \xi_b^{1+\beta}> \zcut^2 \xi_c^{1+\beta}>\zcut^2 \xi_\Lambda^{1+\beta}.
\end{equation}

\section{Results for the $\theta_g$ distribution at low and high transverse momentum}\label{app:theta_g-other-pt}
\begin{figure*} 
\begin{center}
\includegraphics[width=0.49\textwidth]{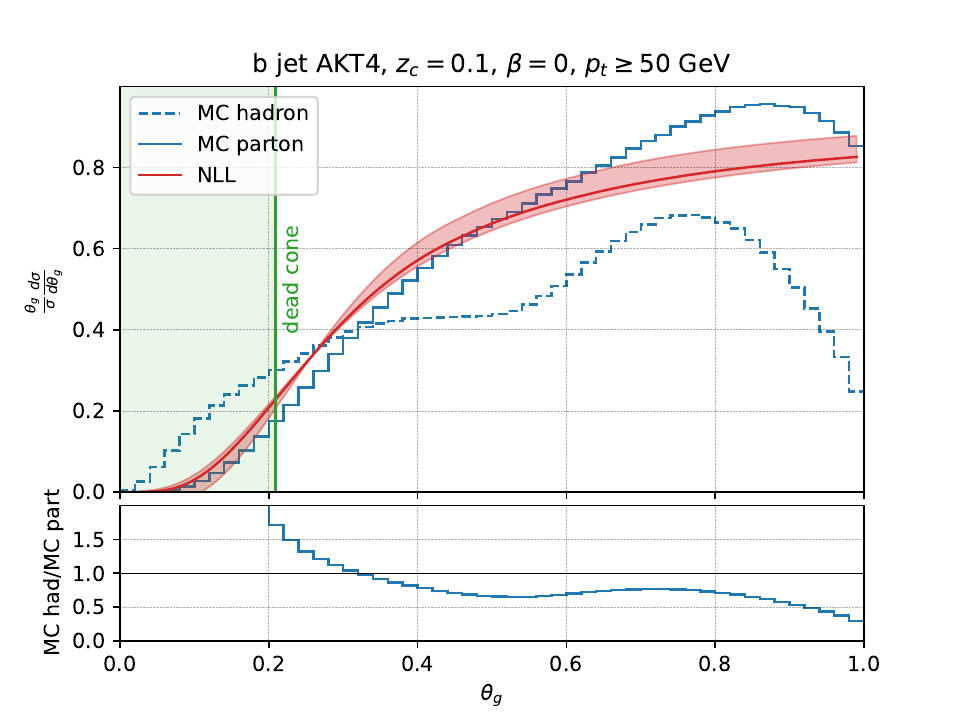}
\includegraphics[width=0.49\textwidth]{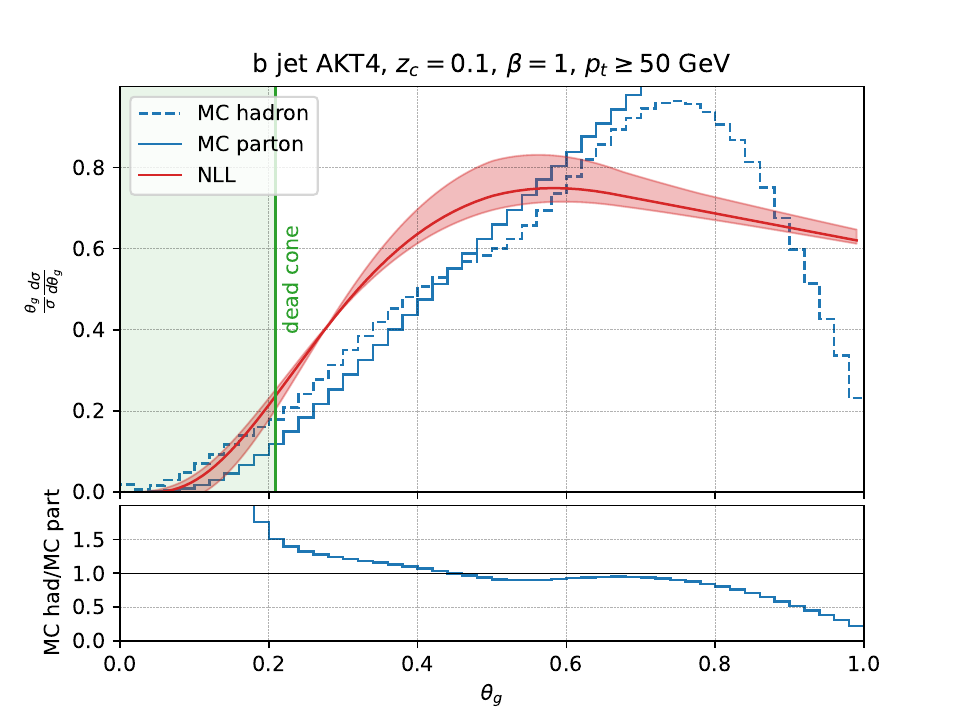}
\includegraphics[width=0.49\textwidth]{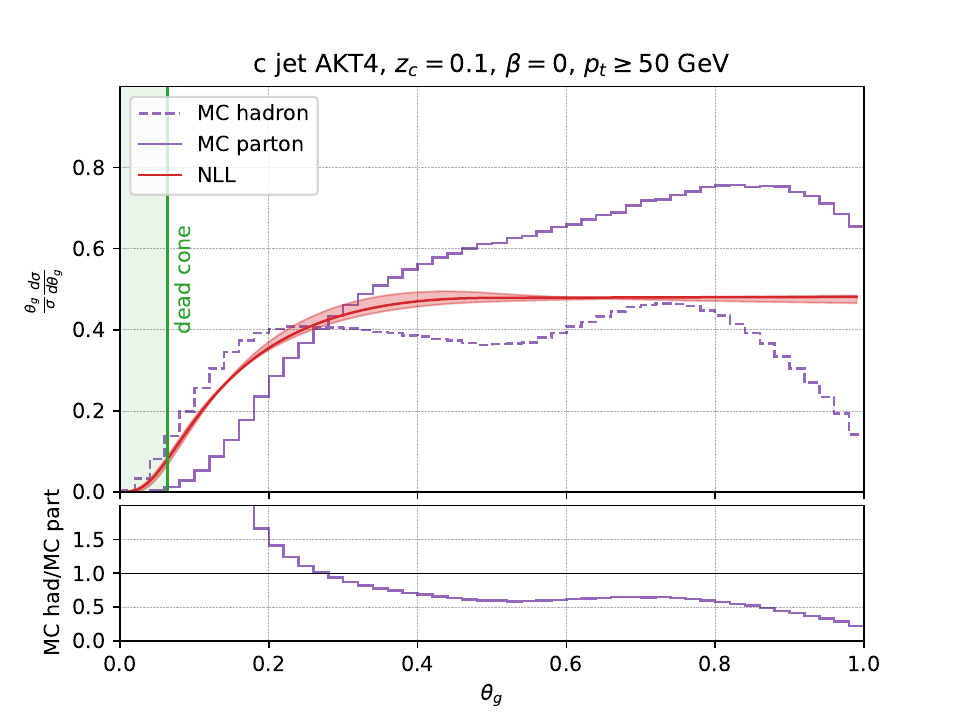}
\includegraphics[width=0.49\textwidth]{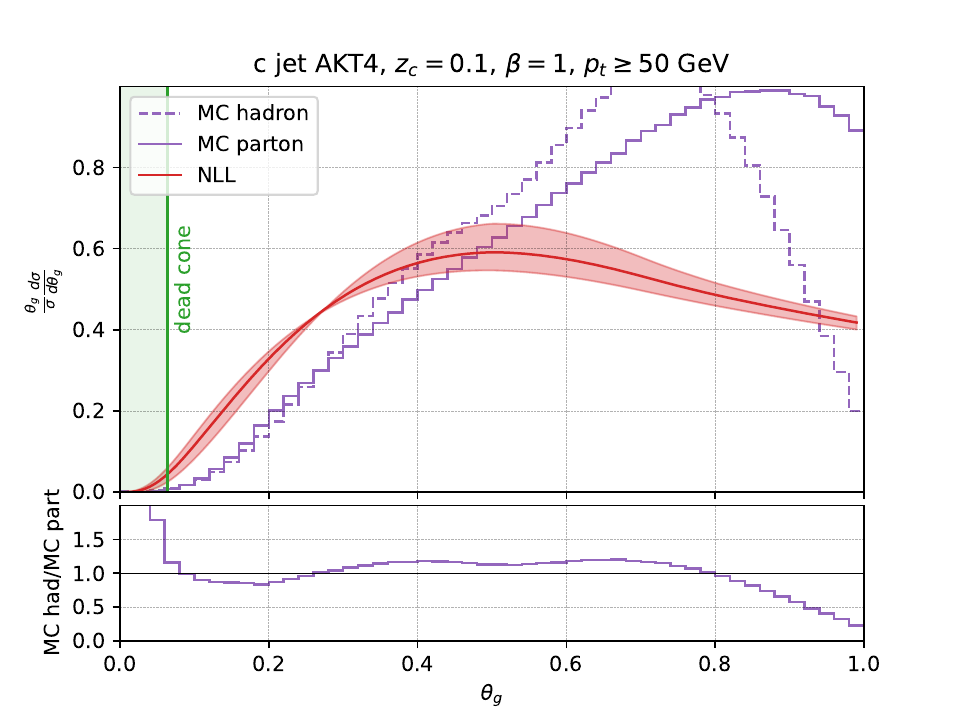}
\includegraphics[width=0.49\textwidth]{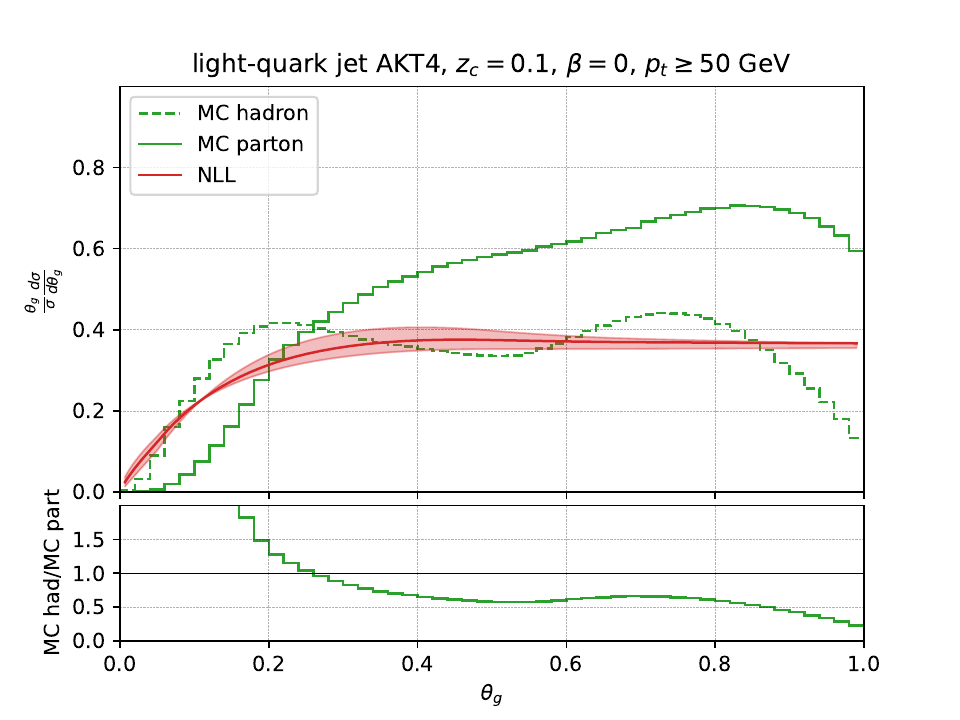}
\includegraphics[width=0.49\textwidth]{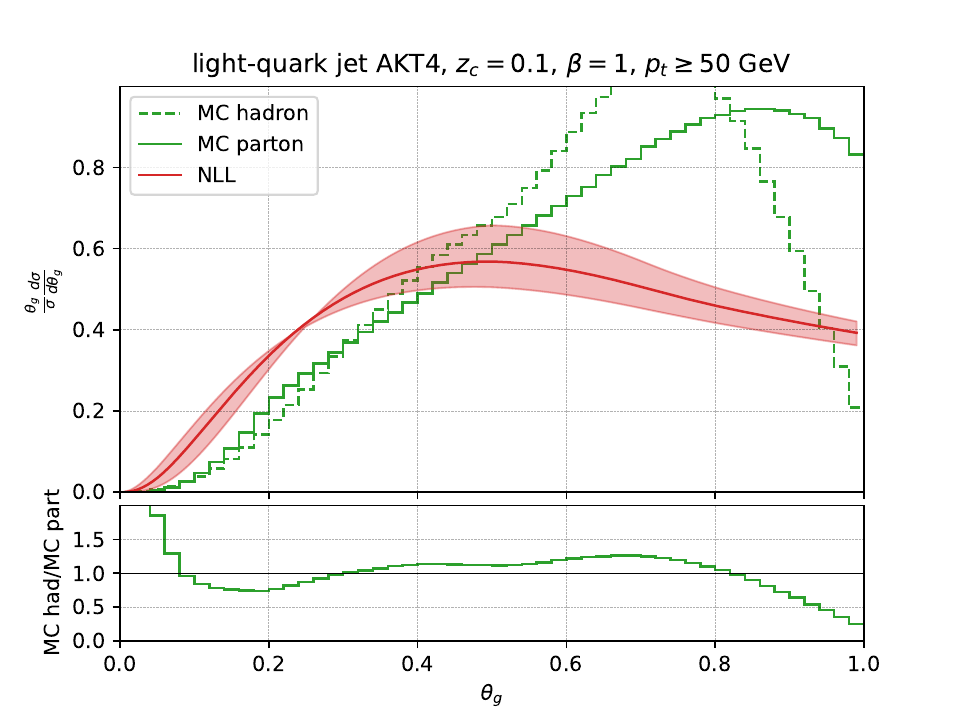}
\end{center}
\caption{Same as Fig.~\ref{fig:thetag}, but for $p_t \ge 50$~GeV.}
\label{fig:thetag-low}
\end{figure*}

\begin{figure*} 
\begin{center}
\includegraphics[width=0.49\textwidth]{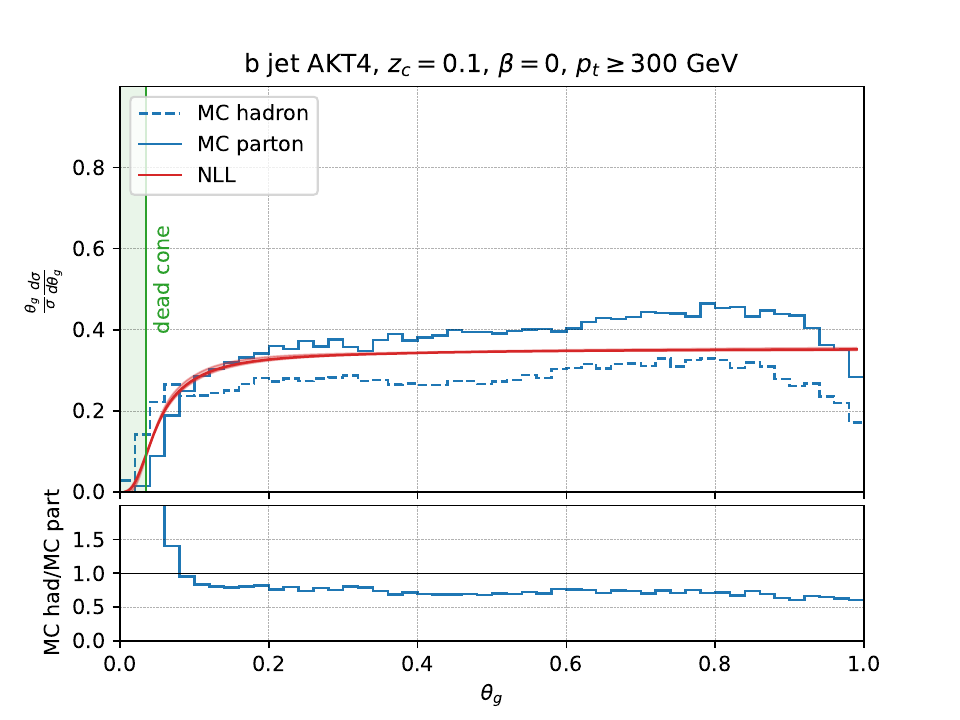}
\includegraphics[width=0.49\textwidth]{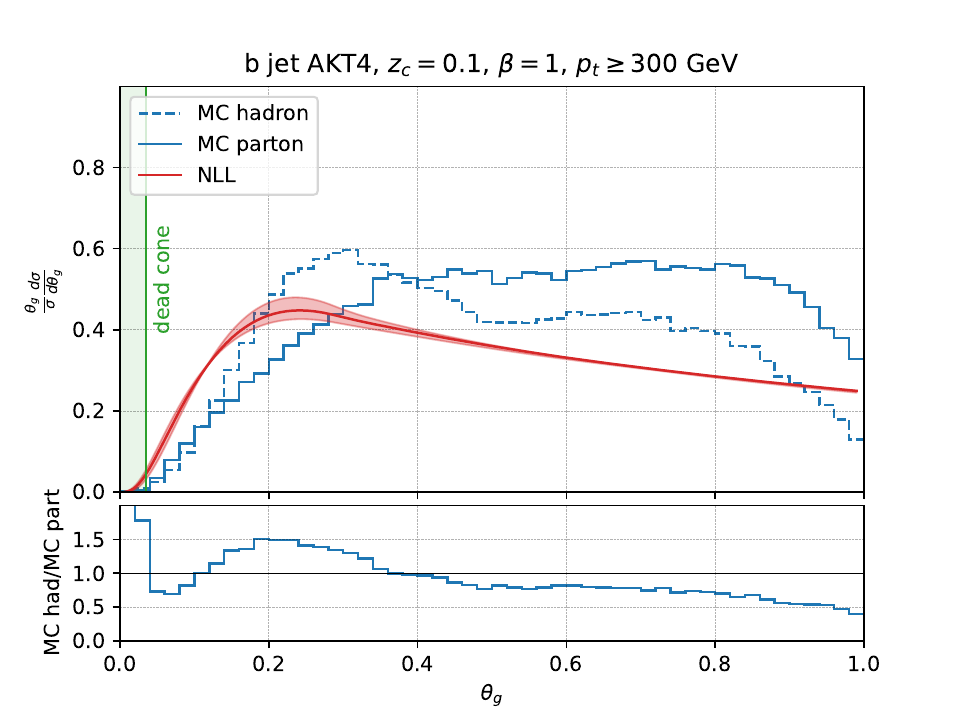}
\includegraphics[width=0.49\textwidth]{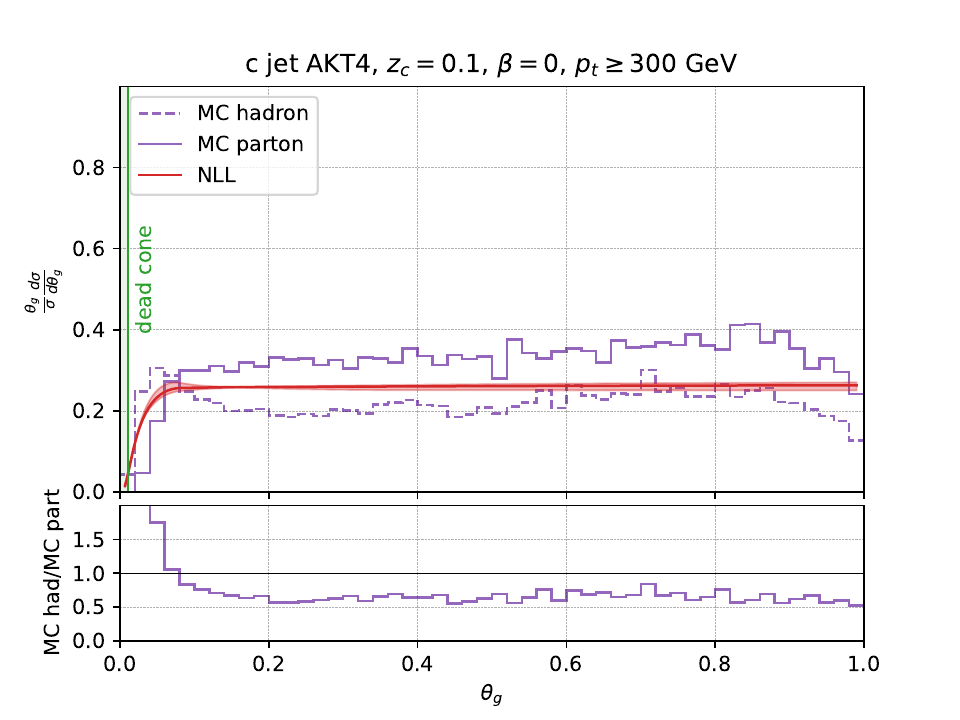}
\includegraphics[width=0.49\textwidth]{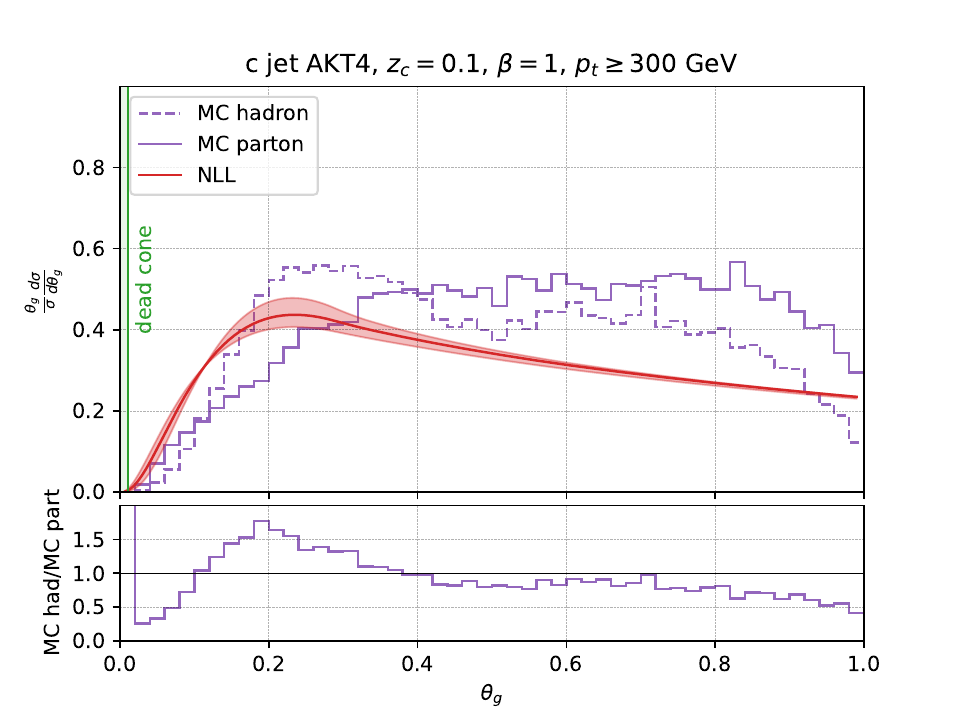}
\includegraphics[width=0.49\textwidth]{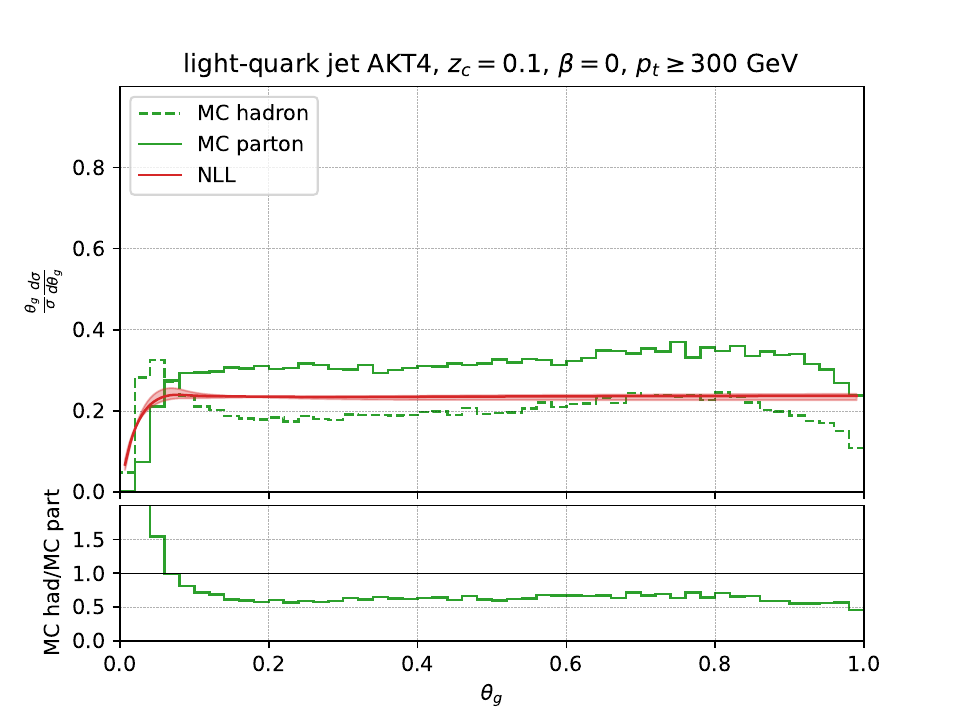}
\includegraphics[width=0.49\textwidth]{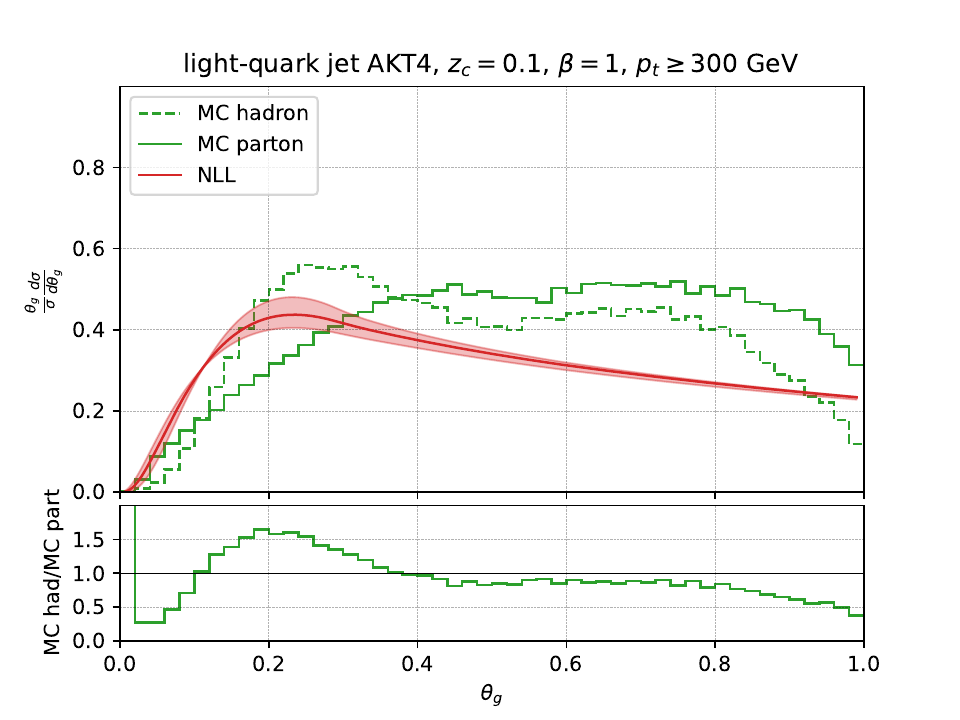}
\end{center}
\caption{Same as Fig.~\ref{fig:thetag}, but for $p_t \ge 300$~GeV.}
\label{fig:thetag-high}
\end{figure*}
We collect here results for the $\theta_g$ distribution at NLL, for two different values of the transverse momentum cut of the jet. 
In Fig.~\ref{fig:thetag-low}, we show the case $p_t\ge 50$~GeV, while in Fig.~\ref{fig:thetag-high} we consider a high-$p_t$ cut, namely $p_t\ge 300$~GeV. Details of the plots are the same as in the main text, Fig.~\ref{fig:thetag}.

The low-$p_t$ selection allows us to expose the dead cone. In the case of $b$-jets, we see that the $\theta_g$ distribution (for $\beta=0$) has clearly a distinct shape compared to the light-quark case.  However, for such low scales $p_t R_0\simeq 20$~GeV, we have to deal with larger non-perturbative effects. This is particularly true for the $\beta=1$ case. 
The high-$p_t$ selection is under better perturbative control, but in this phase-space region, mass effects are essentially negligible. 
This is still interesting because, in this context, heavy-flavour identification provides us with a purified sample of quark jets, i.e. it essentially works as a quark/gluon tagger.

\bibliographystyle{jhep}      

\bibliography{HFSoftDrop/HFSoftDrop}

\end{document}